\documentclass[preprint,5p,times]{elsarticle}

\usepackage{amssymb}
\usepackage{amsmath}
\usepackage[table]{xcolor}
\usepackage{hyperref}
\usepackage{microtype} 
\hypersetup{
    colorlinks=true,        
    linkcolor=black,         
    citecolor=black,         
}
\usepackage{dblfloatfix}
\usepackage{booktabs}
\usepackage{multirow}

\usepackage{marginnote}
\usepackage{soul}
\newcommand{\redhighlight}[1]{%
  \sethlcolor{yellow}%
  \textcolor{red}{\hl{#1}}%
}

\journal{Solar Energy}

\renewcommand{\marginpar}[1]{}
\renewcommand{\marginnote}[1]{}
\renewcommand{\redhighlight}[1]{#1}

\newcommand{\red}[1]{\textcolor{red}{#1}}

\renewcommand{\red}[1]{#1}

\begin{document}

\begin{frontmatter}



\title{Benchmarking Deep Learning Methods for Irradiance Estimation from Sky Images with Applications to Video Prediction-Based Irradiance Nowcasting}



\author[label1]{Lorenzo F. C. Varaschin} 
            
\author[label2]{Danilo Silva} 

\affiliation[label1, label2]{organization={Department of Electrical and Electronic Engineering, Federal University of Santa Catarina},
            addressline={}, 
            city={Florianópolis},
            postcode={}, 
            state={Santa Catarina},
            country={Brazil}}
            
\begin{abstract}

To address the high levels of uncertainty associated with photovoltaic energy, an increasing number of studies focusing on short-term solar forecasting \redhighlight{(i.e. nowcasting)} have been published. Most of these studies use deep-learning-based models to directly forecast a solar irradiance or photovoltaic power value given an input sequence of sky images. Recently, however, advances in generative modeling have led to approaches that divide the \redhighlight{nowcasting} problem into two sub-problems: 1) future event prediction, i.e. generating future sky images; and 2) solar irradiance or photovoltaic power \redhighlight{estimation}, i.e. predicting the concurrent value from a single image. One such approach is the SkyGPT model, whose potential for improvement is shown to be much larger in the \redhighlight{estimation} component than in the generative component. Thus, in this paper, we focus on the solar irradiance estimation problem and conduct an extensive benchmark of deep learning architectures across the widely-used Folsom, SIRTA and NREL datasets. Moreover, we perform ablation experiments on different training configurations and data processing techniques, including the choice of the target variable used for training and adjustments of the timestamp alignment between images and irradiance measurements. In particular, we draw attention to a potential error associated with the sky image timestamps in the Folsom dataset and suggest a possible fix. By leveraging the three datasets, we demonstrate that our findings are consistent across different solar stations. \redhighlight{Finally, we combine our best irradiance estimation model with a video prediction model and obtain state-of-the-art results on the SIRTA dataset.}
\end{abstract}

\begin{keyword}
Solar Irradiance \sep Nowcasting \sep Forecasting \sep Deep Learning \sep Sky Images \sep \redhighlight{Video Prediction}


\end{keyword}

\end{frontmatter}


\section{Introduction}
\label{sec1}


Solar photovoltaic (PV) energy, which harvests the Sun's energy directly into electric energy by using semiconductor cells, has seen an exponential growth in recent years \cite{Kabir.etal.2018.Solar-Energy-Potential}. This has brought forth a new set of challenges to the electrical grid operators \cite{Liu.Bebic.2008.Distribution-System-Voltage, Kakimoto.etal.2011.Voltage-Control-Photovoltaic, Eftekharnejad.etal.2015.Optimal-Generation-Dispatch, Fan.etal.2013.Probabilistic-Power-Flow}, due to the intermittent and uncertain nature of this energy source \cite{Bird.etal.2013.Integrating-Variable-Renewable, Shah.etal.2015.Review-Key-Power}. The primary cause of uncertainty in PV power generation is the presence of clouds, which can abruptly increase or decrease the amount of solar radiation harnessed by PV panels at any given time. Consequently, a large number of studies have been published in recent years with the goal of creating accurate solar forecasting models. 


In the very short term solar forecasting ($\leq$ 60 minutes forecasting horizon), \redhighlight{commonly referred to as solar nowcasting}, all-sky image (ASI) based models have achieved state-of-the-art results, due to their ability to capture the cloud information with a very high spatio-temporal resolution \cite{Rajagukguk.etal.2020.Review-Deep-Learning}. These images are typically combined with convolutional neural networks (CNN), which extract features from the images that will be used to predict the solar value. The following paragraphs briefly describe some of these approaches.

The authors of \cite{Wen.etal.2021.Deep-Learning-Based} use a ResNet18 architecture \cite{He.etal.2015.Deep-Residual-Learning} to encode the image features, which are then used to predict the solar irradiance 5 to 10 minutes ahead. However, instead of feeding a static image to the model, they stack multiple images at different timestamps and gain up to a 7\% performance boost. The method proposed in \cite{Yang.etal.2021.3D-CNN-Based-Sky-Image} used a 3D CNN \cite{Tran.etal.2015.Learning-Spatiotemporal-Features} to forecast the clear sky index $k_t$, which results in a 15.2\% improvement over the smart persistence model. In \cite{Paletta.etal.2021.Benchmarking-Deep-Learning}, a benchmark study of four different architectures (CNN, CNN+LSTM, 3D-CNN and ConvLSTM) is conducted and, while the more complex models tend to perform better, it is shown that they all share a fundamental problem: \redhighlight{the predicted future irradiance values are often delayed relative to the ground truth measurements, resulting in irradiance ramp predictions that are systematically delayed relative to the ground truth.}


A multi-modal transformer based framework is proposed in \cite{Liu.etal.2023.Transformer-based-Multimodal-learning-Framework}, where the solar irradiance time series is encoded by a transformer \cite{Vaswani.etal.2023.Attention-All-You}, and the sky images are transformed into optical flow maps and encoded by a vision transformer \cite{Dosovitskiy.etal.2021.Image-Worth-16x16}, which are then fused with a cross modality attention block to predict the solar irradiance up to 30 minutes ahead. In \cite{Feng.Zhang.2020.SolarNet-Sky-Image-based}, a VGG16 \cite{Simonyan.Zisserman.2015.Very-Deep-Convolutional} inspired model is developed specifically for the solar irradiance \redhighlight{nowcasting} task, the SolarNet model, which outperformed several of their benchmark models. In \cite{Sun.etal.2019.Short-term-Solar-Power-1}, different input and output configurations are explored for the Sunset model, which is another model specifically designed for the solar \redhighlight{nowcasting} task, but focuses on the solar PV power instead of solar irradiance. 

The methodology proposed in \cite{Nie.etal.2024.SkyGPT-Probabilistic-Ultra-short-term} is particularly interesting as it separates the future event prediction task from the irradiance \redhighlight{estimation} task. This is achieved by using a sky image generation model, SkyGPT, which is trained auto-regressively to predict future sky images based on past image sequences. With the generated future sky image, a modified version of the U-Net architecture \cite{Ronneberger.etal.2015.U-Net-Convolutional-Networks} is then used to \redhighlight{estimate} the concurrent PV power. By decoupling the future event prediction task (SkyGPT) from the PV power \redhighlight{estimation} task (U-Net), they demonstrate that improving the SkyGPT model (such that it could perfectly predict the future sky image) would only boost the performance by 13\%, whereas improving the \redhighlight{estimation} model could yield up to a 64\% performance boost (see Figure 7 of \cite{Nie.etal.2024.SkyGPT-Probabilistic-Ultra-short-term}).

Inspired by the larger potential improvement, this study focuses primarily on the solar irradiance \redhighlight{estimation} task, using exclusively sky images as input, i.e. given an image at timestamp $t$, we want to predict the corresponding irradiance value at the same timestamp $t$. Although this is different from PV power \redhighlight{estimation}, we expect the results to generalize well across both tasks, since they share the same underlying nature and are highly correlated.



Compared to solar \redhighlight{nowcasting}, fewer studies have tackled the solar \redhighlight{estimation} task. In \cite{Papatheofanous.etal.2022.Deep-Learning-Based-Image} they perform solar irradiance \redhighlight{estimation} and show that appending a Sun mask as a fourth channel to the images significantly increases the \redhighlight{estimation} performance, which we also investigate in this study. In \cite{Nie.etal.2023.SKIPPD-SKy-Images} they adapt the Sunset model for the solar PV power \redhighlight{estimation} task, which we use in this study for irradiance \redhighlight{estimation}.

We begin our study by conducting an extensive \redhighlight{irradiance estimation} 
benchmark of ten deep learning architectures across the Folsom, SIRTA and NREL datasets, which are three datasets commonly used in the solar forecasting literature. We then perform several ablation experiments, including the choice of the target variable used for training and adjusting the timestamp alignment between the images and irradiance measurements. \redhighlight{Finally, we train a video prediction model and combine it with our irradiance estimation model to conduct proof of concept experiments on the two-step approach in solar irradiance nowcasting. All of our experiments are evaluated using the root mean squared error (RMSE) and mean absolute error (MAE) and we show that our findings are consistent across all three datasets. Our main contributions can be summarized as follows:}


\begin{enumerate}[1.]
\item We conduct an extensive \redhighlight{irradiance estimation} benchmark of deep learning architectures across three solar irradiance datasets and find that the ResNet50 architecture provides significant gains over the U-Net architecture used in \cite{Nie.etal.2024.SkyGPT-Probabilistic-Ultra-short-term};
\item We demonstrate that appending a Sun mask as a fourth channel to the images as in \cite{Papatheofanous.etal.2022.Deep-Learning-Based-Image} improves the performance when the model's target variable is the unnormalized solar irradiance value, but is detrimental otherwise;
\item We demonstrate that the Folsom dataset contains an error associated with their sky image timestamps and a potential fix is investigated;
\item We demonstrate that shifting the irradiance measurements of the training set by a few seconds in time can significantly improve the test set performance.
\item \redhighlight{With our best irradiance estimation model and a simple video prediction model, we obtain state of art nowcasting results on the SIRTA dataset.}
\end{enumerate}

The rest of this article is structured as follows: \autoref{irr_prelim} briefly describes the fundamental concepts of solar irradiance, such as the different irradiance components and clear sky models; \autoref{data_desc} describes the three datasets that were used in this study, as well as the different processing steps that were applied; \autoref{methods} describes our methods, where we go into detail about our trained models and evaluation metrics; \autoref{results} details all the results and experiments conducted in this study; \autoref{conclusion} concludes the article.

\section{Solar Irradiance Fundamentals}
\label{irr_prelim}
\subsection{Irradiance Components}
\label{subsec1}

Solar irradiance is a measurement of the rate of solar radiation energy that arrives at a surface area and is typically measured in $\text{W/m}^2$. The mean solar irradiance at the top of the Earth's atmosphere, also known as the solar constant \cite{..Glossary-Solar-Radiation}, is approximately 1366 $\text{W/m}^2$. After entering the Earth's atmosphere, the solar radiation will be attenuated by the complex interactions with the atmospheric constituents, resulting in two irradiance components at the ground level: diffuse horizontal irradiance (DHI) and direct normal irradiance (DNI). \redhighlight{The DHI is the amount of solar radiation incident upon a horizontal plane that has been scattered or reflected by atmospheric constituents or nearby objects.} The DNI measures the amount of solar radiation that is coming directly from the direction of the Sun beam. The sum of these two components on a horizontal plane results in the global horizontal irradiance (GHI), denoted by:
\begin{equation}
\label{eq: ghi_eq}
\text{GHI} = \text{DNI}\times \cos{\theta_z} + \text{DHI}
\end{equation}
where $\theta_z$ is the \redhighlight{Solar Zenith Angle (SZA)}. This component is the primary focus of this study.

\subsection{Clear Sky Models}
Predicting the GHI under clear sky conditions is a well-defined problem, and there are hundreds of different clear sky models available in the literature, with varying but generally excellent performances across many different climates \cite{Antonanzas-Torres.etal.2019.Clear-Sky-Solar, Sun.etal.2019.Worldwide-Performance-Assessment}. Clear sky models can also be used to obtain the clear sky index $k_t$, that normalizes the GHI measurement according to:
\begin{equation}
\label{eq: clear_sky_index_eq}
k_t = \frac{I}{I^\text{clr}}
\end{equation}
where $I$ is the GHI measurement and $I^\text{clr}$ is the clear sky GHI predicted by a clear sky model. In this study, the Simplified Solis Clear Sky Model \cite{Ineichen.2008.Broadband-Simplified-Version} is used.

Intuitively, $I^\text{clr}$ would represent the highest possible GHI value and $k_t$ should range between 0 (overcast) to 1 (clear sky). However, $k_t$ can assume values greater than 1, not only because of the imperfections of the clear sky model, but mainly due to a phenomenon called \textit{overirradiance} \cite{DoNascimento.etal.2019.Extreme-Solar-Overirradiance}, where the presence of clouds actually increases the GHI through solar beam reflection.

Other than normalization, the $k_t$ index and the clear sky prediction $I^\text{clr}$ can be used to obtain the smart persistence model (SPM), which is a naive model that assumes that the current $k_t$ will stay the same throughout the entire forecasting horizon $h$: 
\begin{equation}
\label{eq: smart_persistence_model}
I_{t+h} = k_t \times I^{\text{clr}}_{t+h}
\end{equation}
This model is commonly used as a baseline reference for solar forecasting models.  

Another commonly used index is the clearness index $K_t$ \cite{Inman.etal.2013.Solar-Forecasting-Methods}, which uses the extraterrestrial irradiance instead to normalize the GHI:
\begin{equation}
\label{eq: clearness_sky_index_eq}
K_t = \frac{I}{I^\text{extr}}
\end{equation}
where $I^\text{extr}$ is the extraterrestrial GHI. The main advantage of this index is that $I^\text{extr}$ is much easier to model than $I^\text{clr}$, and can be written as:

\begin{equation}
I^\text{extr} = I_0\times \cos{\theta_z}
\end{equation}
where $I_0$ is the solar constant and $\theta_z$ is the SZA. The choice between both normalization indices is investigated in \autoref{model_target}.

\subsection{Detection of Clear Sky Periods}
\label{clear_sky_detection}
Although sky condition classification is outside the scope of this study, detecting clear/cloudy sky periods can still be used to split the testing set into two smaller subsets, thus providing a more detailed and insightful evaluation of the model's performance under varying sky conditions. Based on a recent review of 21 different clear sky identification methods \cite{Gueymard.etal.2019.Posteriori-Clear-sky-Identification}, and considering the fact that the algorithm is readily available through Python's pvlib library, we chose the Reno \cite{Reno.Hansen.2016.Identification-Periods-Clear} method. This is a very simple and effective algorithm that only requires the measured GHI and the expected clear sky GHI data (from the Simplified Solis model, in our case) as inputs. By comparing 5 different statistics of the measured and expected clear sky values over a sliding time window (typically 10 minutes), the algorithm flags a measurement as either clear or cloudy. 

\begin{figure}
\centering
\includegraphics[width=\linewidth]{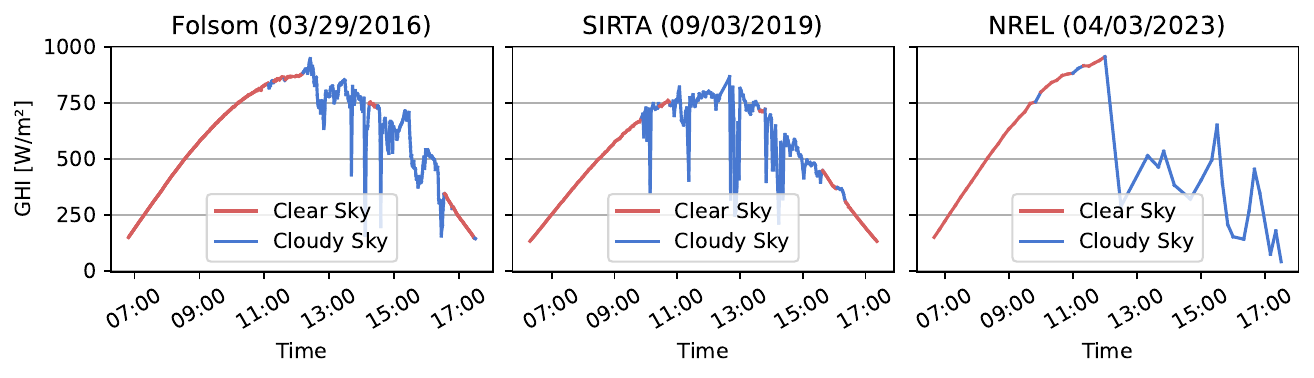}
\caption{\redhighlight{Clear and cloudy sky measurements in each dataset obtained through the Reno algorithm} \cite{Reno.Hansen.2016.Identification-Periods-Clear}.}\label{im: cs_instants}
\end{figure}

\autoref{im: cs_instants} shows an example of this method being applied. From our testing, this method works very well for all three datasets in the majority of cases. For more details on this algorithm, we refer the reader to the original paper \cite{Reno.Hansen.2016.Identification-Periods-Clear}.

\section{Data Description and Processing}
\label{data_desc}
In this section, we describe the raw data used in this study, obtained from three different datasets, as well as the preprocessing pipeline that is made prior to training the models. 

\subsection{Folsom Dataset} \label{folsom_dataset_descp}
The first dataset used in this study is the Folsom dataset ($\mathcal{D}^{\text{folsom}}$) \cite{Pedro.etal.2019.Comprehensive-Dataset-Accelerated}, collected in Folsom, California (38.642$^\circ$ N and 121.148$^\circ$ W), over a 3-year period (2014-2016). This dataset has been widely used by the solar forecasting community \cite{Papatheofanous.etal.2022.Deep-Learning-Based-Image, Wen.etal.2021.Deep-Learning-Based, Yang.etal.2020.Probabilistic-Solar-Forecasting, Song.etal.2024.Intra-hour-Solar-Irradiance, Wang.etal.2023.Hybrid-Ensemble-Learning, Ruan.etal.2024.Use-Sky-Images, Zhang.etal.2023.Advanced-Multimodal-Fusion, NunesMaciel.etal.2024.Hybrid-Prediction-Method} and it contains a variety of irradiance correlated data, such as sky images, temperature, wind, humidity, among others. In this benchmark study, only the sky images and irradiance data were used.

The sky images were captured during the daylight by a fish-eye lens camera with a resolution of $1536\times 1536$ pixels at a 1-minute sampling interval, totaling over 750,000 images. The GHI and DHI data are collected throughout the entire day by a second generation RSR, which contains two Licor LI-200SZ pyranometers, and the DNI is computed from the measured GHI, DHI and the solar zenith angle $\theta_z$, as per \autoref{eq: ghi_eq}. While all three irradiance data are measured at a 1-second sampling interval, only their 1-minute averages are provided.

Examples of the sky images are shown in the leftmost column of \autoref{im: image_data} and, unlike the other two datasets, \redhighlight{the camera exposure time is the same for all images in $\mathcal{D}^{\text{folsom}}$, though the exact value of this setting is unavailable in} \cite{Pedro.etal.2019.Comprehensive-Dataset-Accelerated} \redhighlight{and the image metadata. Also unlike the other datasets, $\mathcal{D}^{\text{folsom}}$ exhibits a few instances of lens contamination due to dust, dirt, and condensation. Although we believe that automatic mitigation strategies for this issue are beyond the scope of this study, we think it is important to highlight it and to inform future research, since this can have an impact on model performance. Thus, we manually identified three representative examples in} \autoref{im: dirty_samples}. The GHI distribution of $\mathcal{D}^{\text{folsom}}$ is shown in the left panel of \autoref{im: irradiance_data}, where it is clear that the density of large irradiance values in this dataset is much higher, i.e. it has a larger number of clear sky \redhighlight{instances} than the other two datasets.

\subsection{SIRTA Dataset}
The second dataset we explored was the SIRTA dataset ($\mathcal{D}^{\text{sirta}}$) \cite{Haeffelin.etal.2005.SIRTA-Ground-based-Atmospheric}, which was collected in Palaiseau, France (48.713$^\circ$ N and 2.208$^\circ$ E) at the Site Instrumental de Recherche par Télédétection Atmosphérique (SIRTA) observatory. The SIRTA observatory has been rigorously collecting high quality irradiance, sky image and meteorological data for more than 10 years and their dataset is also extensively used by the community \cite{Paletta.etal.2021.Benchmarking-Deep-Learning, Paletta.etal.2022.ECLIPSE-Envisioning-CLoud, Jain.etal.2024.Holistic-Lightweight-Approach, Paletta.etal.2024.Improving-Cross-site-Generalisability, Nie.etal.2024.Sky-Image-based-Solar, Insaf.etal.2021.Global-Horizontal-Irradiance, AlAsmar.etal.2021.Improvement-Solar-Irradiance}. For our benchmark study, we only select the 2017-2019 data, as this seems to be the most common choice across the literature.

The sky images (middle images of \autoref{im: image_data}) were captured with a $768\times 1024$ resolution  at a 1-minute sampling interval in 2017 and 2-minute in 2018 and 2019 and come in two forms: 1) a long exposure form that shows the sky in more detail but can have saturated pixels in the area around the Sun \redhighlight{(1/1000 s exposure)}; 2) a short exposure form that shows the area around the Sun less saturated and in more detail, but the rest of the sky is shown in less detail \redhighlight{(1/2000 s exposure)}. The GHI and DHI are measured by a CMP22 pyranometer and the DNI is measured by a CHP1 pyrheliometer, all at a 1-second sampling interval, but only the 1-minute averages are provided.

Compared to the other two datasets, the GHI distribution in $\mathcal{D}^{\text{sirta}}$ (middle panel of \autoref{im: irradiance_data}) is much more concentrated at lower values, i.e. this dataset has much more cloudy/overcast samples which, generally speaking, are much harder for the models to learn than clear sky samples.

\subsection{NREL Dataset}
Lastly, we also analyzed the National Renewable Energy Laboratory (NREL) dataset ($\mathcal{D}^{\text{nrel}}$) \cite{Stoffel.Andreas.1981.NREL-Solar-Radiation}, collected at NREL's Solar Radiation Research Laboratory (SRRL), located in Golden, Colorado (39.742$^\circ$ N and 105.18$^\circ$ W). The SSRL has been measuring irradiance correlated data since 1981 and, once again, many studies have been published using this dataset: \cite{Jonathan.etal.2024.Radiant-Shift-Attention-embedded, Gao.Liu.2022.Short-term-Solar-Irradiance, Liu.etal.2023.Transformer-based-Multimodal-learning-Framework, Zuo.etal.2022.Ten-minute-Prediction-Solar, Feng.etal.2022.Convolutional-Neural-Networks}. For our benchmark studies, we only selected the sky images and irradiance data collected from 2020 to 2023. \redhighlight{Unlike the other datasets, $\mathcal{D}^{\text{nrel}}$ provides multiple independent measurements of the same irradiance component. We used the following sensors:}


\begin{enumerate}[a.]
\item \redhighlight{DNI: CHP1-1, CHP1-2, sNIP, NIP and DR02 pyrheliometers;}
\item \redhighlight{DHI: CM22-2, CM22-1, SR25, Research 1 and 8-48 pyranometers;}
\item \redhighlight{GHI: CMP22, CMP22-1, CMP22-2, CMP11 and CM6b pyranometers.}
\end{enumerate}

\redhighlight{By leveraging all of these sensors, we were able to detect and filter out several faulty measurements. To do so, we implement the following: let $M(X)$ denote the median of a set of measurements $X$. We first compute the medians of the DHI and DNI measurements, $M(\text{DHI})$ and $M(\text{DNI})$, and use them to estimate the global component, $\text{GHI}_{\text{calculated}}$, according to} \autoref{eq: ghi_eq}. \redhighlight{We then compare this calculated value with the median of the GHI measurements, $M(\text{GHI})$, and keep only the samples that satisfy:}

\begin{equation}
    |\text{GHI}_{\text{calculated}}-M(\text{GHI})| \leq 10\text{ W/m}^2
\end{equation}

\redhighlight{This filtering step retains 961,069 of the original 1,063,282 measurements and improves the overall reliability of the irradiance labels.} The GHI distribution is plotted in the right panel of \autoref{im: irradiance_data}. 

\begin{figure}
\centering
\includegraphics[width=\linewidth]{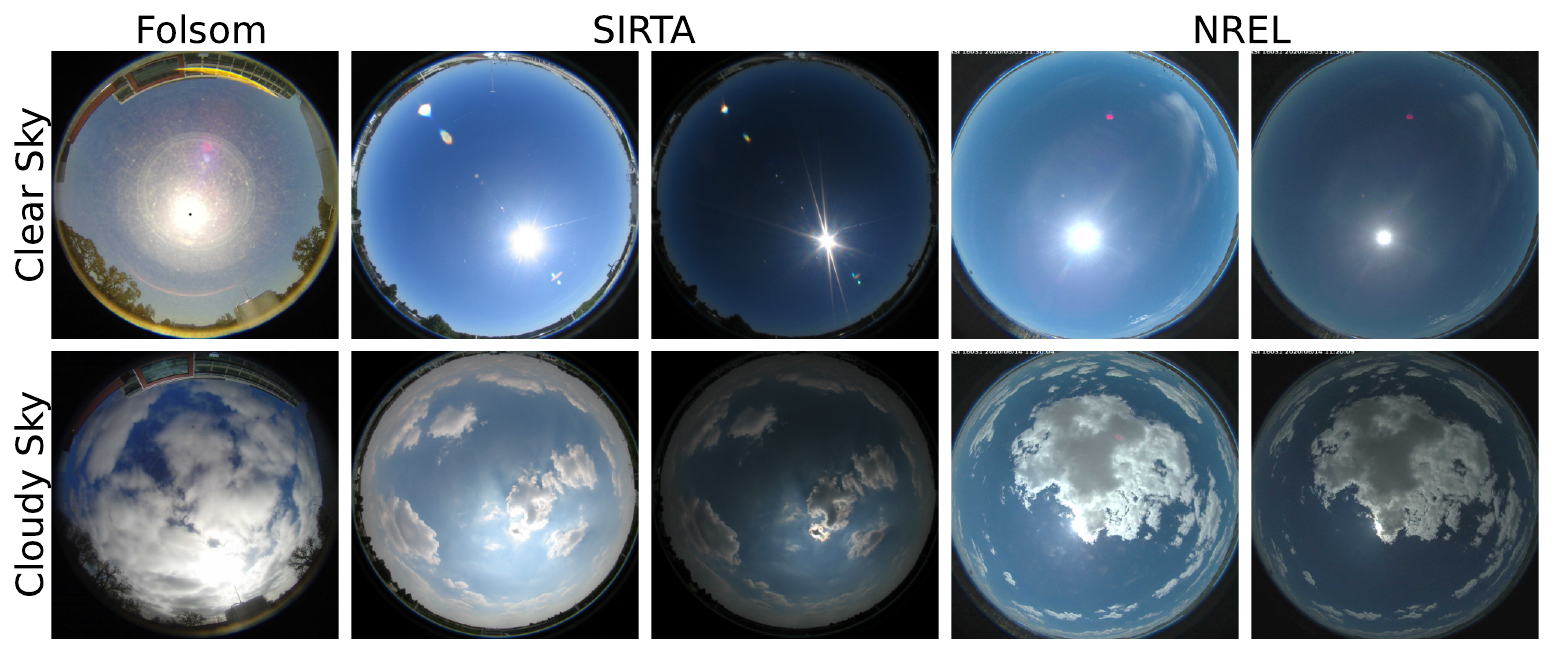}
\caption{\redhighlight{Clear and cloudy sky image examples for each dataset used in this study. $\mathcal{D}^{\text{sirta}}$ and  $\mathcal{D}^{\text{nrel}}$ also provide a short exposure alternative for each image, so that the area around the Sun can be shown in greater detail. The depicted images were cropped such that they have the same shape.}}\label{im: image_data}
\end{figure}

The sky images are provided at a $1536\times 1536$ pixel resolution in both long and short exposure times (rightmost images of \autoref{im: image_data}), \redhighlight{though the exact value of this setting is also unavailable in this dataset}. The sky images are only available once every 10 minutes, which makes this a much smaller dataset compared to the other two (for our selected years in each dataset). The impact that a smaller training dataset has on the model performance is further discussed in \autoref{sample_interval}

\begin{figure}
\centering
\includegraphics[width=\linewidth]{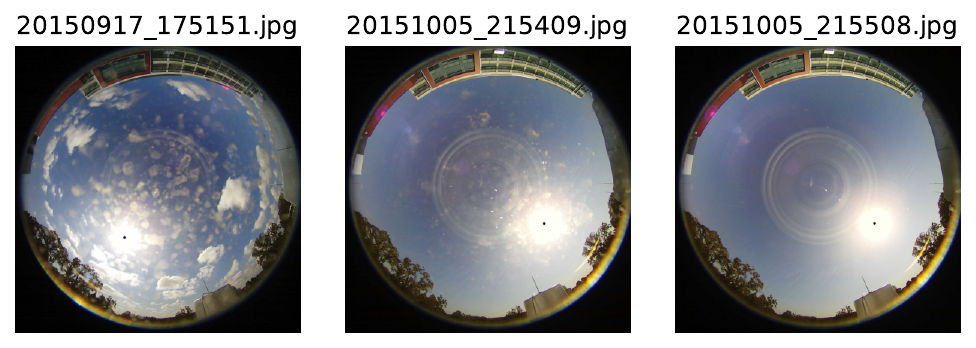}
\caption{\redhighlight{Lens contamination examples in $\mathcal{D}^{\text{folsom}}$: an image with severe contamination (left), an image taken immediately before lens cleaning (middle), and one taken immediately after cleaning (right).}} \label{im: dirty_samples}
\end{figure}

\subsection{Sky Image Data Processing}
The image processing was essentially the same for all 3 datasets. First we compute the mean image of each dataset (using only the long exposure images for $\mathcal{D}^{\text{sirta}}$ and $\mathcal{D}^{\text{nrel}}$) and then define a circular region of interest (ROI), with radius $R$ and center $C$, which is used as a binary mask for that dataset (\redhighlight{see} \autoref{tab:roi_mask_params}). These masks are applied to all the images in their respective datasets to \redhighlight{erase} every pixel outside their ROI. \redhighlight{This is done with the intention of standardizing the image data across all datasets, such as removing the timestamp annotation in the upper left corner of the $\mathcal{D}^{\text{nrel}}$ images and removing any lens flare and/or noise effect in the black area of the images.} Following this, we center square crop the images on $C$ with length $2 R$. Finally, the images are downscaled to a $64\times 64$ resolution using the Python's PIL library. To avoid any potential JPEG compression losses \cite{Parmar.etal.2022.Aliased-Resizing-Surprising}, the resized images are saved to disk in \textit{.npy} format.

\begin{figure}
\centering
\includegraphics[width=\linewidth]{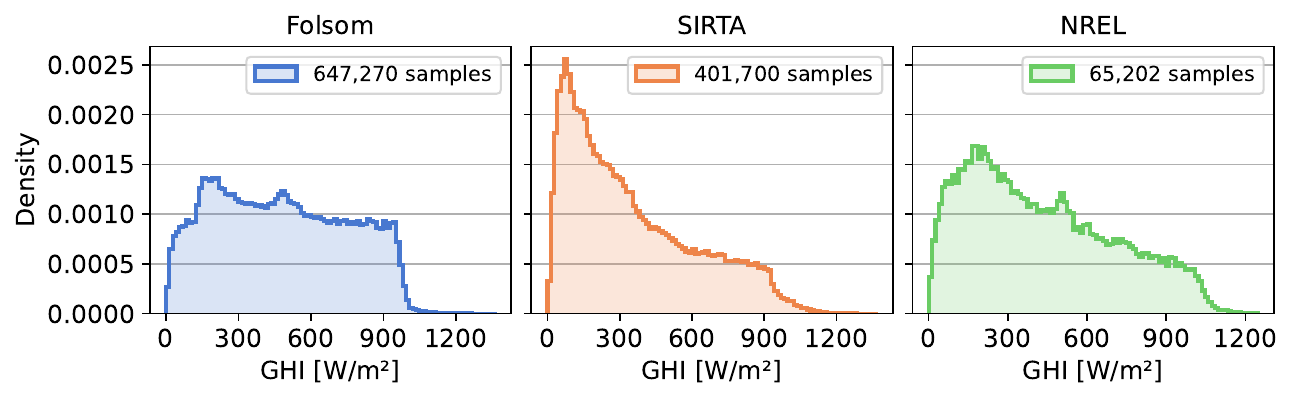}
\caption{The GHI data distributions for the three datasets used in this study, plotted after the data processing pipeline. This distribution is based on the total number of image/irradiance pairs, not the total number of irradiance measurements.}\label{im: irradiance_data}
\end{figure}

\begin{table}
  \centering
  \caption{\redhighlight{Circular region of interest (ROI) radius and center values used in each dataset.}}
  \label{tab:roi_mask_params}
      \begin{tabular}{ccc}  
      \toprule
      Dataset & $R$ (pixels) & $C$ (x, y) \\
      \toprule
      Folsom & 755 & (768, 768) \\
      \cline{1-3}
      SIRTA & 340 & (502, 394)\\
      \cline{1-3}
      NREL & 755 & (768, 768)\\
      \cline{1-3}
      \bottomrule
      \end{tabular}
\end{table}

\subsection{Irradiance Data Processing}
\label{irr_data_proc}
The irradiance data processing was also largely the same for all 3 datasets. First, we resample the data to a 1-second frequency through linear interpolation (more details on why this is done in the following section). To avoid large interpolation errors, this is only done for the samples that are 1-minute apart, i.e. periods with 2 or more minutes of missing data are discarded. Following the linear interpolation, we remove all samples where $\theta_z > 80^\circ$, as these samples provide little information \redhighlight{for our application.}

\begin{figure}
\centering
\includegraphics[width=0.8\linewidth]{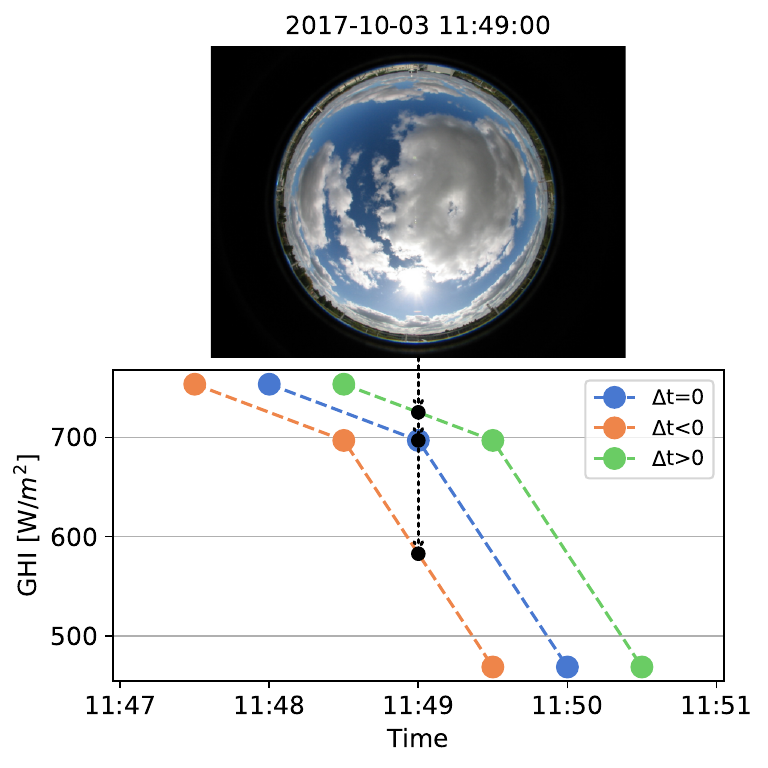}
\caption{\redhighlight{The effect of applying a $\Delta \text{t}$ to the irradiance time series. A negative $\Delta \text{t}$ essentially advances the samples in time, by shifting the curve to the left, whereas a positive $\Delta \text{t}$ has the opposite effect, by shifting it to the right.}} \label{im: timedelta_exmple}
\end{figure}

The next processing step is \textbf{only applied to the training set} of each dataset (see \autoref{train_test_val_splits} for the train/test splits) and consists of shifting the irradiance measurements in time by applying a time offset ($\Delta \text{t}$), which we do according to: 
\begin{equation}
\label{timedelta_eq}
T_{\text{new}} = T_{\text{orig.}} + \Delta \text{t}
\end{equation}
where $T_{\text{new}}$ and $T_{\text{orig.}}$ are the new and original irradiance measurement timestamps, respectively. If a $\Delta \text{t} < 0$ is applied, then we are advancing the irradiance series in time, relative to the sky image series, whereas a $\Delta \text{t} > 0$ would have the opposite effect, as shown in \autoref{im: timedelta_exmple}. Depending on how these irradiance values were averaged, we hypothesize that this can be beneficial. For instance, if the value of $I$ at 14:00:00 represents the average GHI values from 13:59:01 to 14:00:00 (backward average), then a $\Delta \text{t} < 0$ could diminish the delay introduced by such an average. \redhighlight{Thus, we shift the irradiance measurements by -20s, 10s and 30s in $\mathcal{D}^{\text{folsom}}$, $\mathcal{D}^{\text{sirta}}$ and $\mathcal{D}^{\text{nrel}}$, respectively, but also investigate other time offsets in} \autoref{timedelta}. \redhighlight{The reason why a $\Delta \text{t}$ is not applied to the testing set is because doing so would lead to unfair comparisons among the nowcasting results. For example, if $\Delta \text{t} > 0$ (e.g. 30s) is applied to the testing set, then, given the same ASI context, the evaluation of a 10 minute lead time prediction in the shifted time series would be equivalent to a 9:30 minutes lead time in the unshifted time series, which is fundamentally easier to predict. Similarly using $\Delta \text{t} < 0$ (e.g. -30s) would be equivalent to a 10:30 minute lead time, which is fundamentally harder.}

\subsection{Sky Image and Irradiance Alignment}
\label{image_alignment}
\begin{figure*}
\centering
\includegraphics[width=\linewidth]{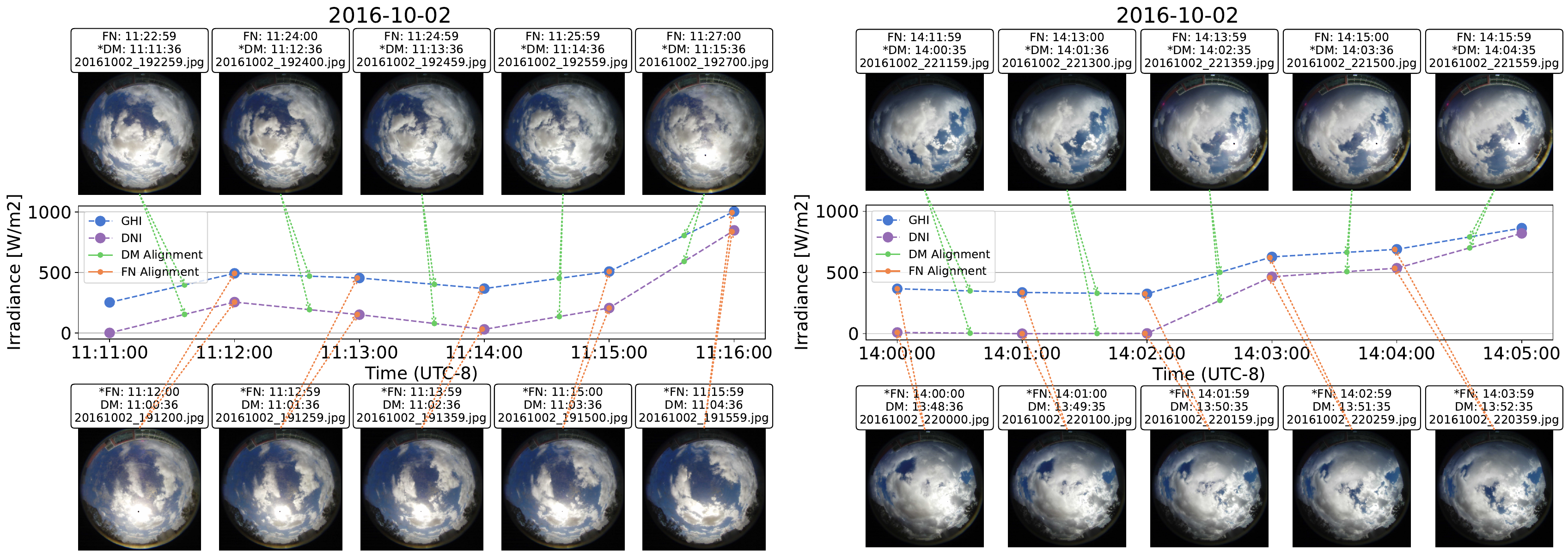}
\caption{\redhighlight{Two distinct 5-minute GHI and DNI windows in $\mathcal{D}^{\text{folsom}}$ which show the corresponding \red{Date Modified (DM)} aligned images in the top row and the \red{File Name (FN)} aligned images in the bottom row. The middle row shows the linearly interpolated GHI data, plotted in UTC-8 timezone\red{. The titles above each image indicate their corresponding FN and DM timestamps (converted to UTC-8 timezone), as well as their JPEG file names. The * marker indicates the timestamp used to align the image to the irradiance plot.}}}\marginnote{\red{Rev.4 Com.1}}\label{im: date_modif1}
\end{figure*}

Our sky image and irradiance alignment for each dataset was very different, especially for $\mathcal{D}^{\text{folsom}}$. Initially, we aligned the sky images with the irradiance data based on the name of the image files, i.e. an image sample $x^{\text{folsom}}_t$, with file name timestamp $t$, was labeled with the irradiance sample $I^{\text{folsom}}_t$. \redhighlight{While this seems like the most natural choice, we noticed, during our initial experiments, that the model's largest errors were samples were the prediction actually seemed quite reasonable, since it predicted larger GHI values when the Sun was clearly visible and lower values when it was not. This raised the suspicion that the labels might be misaligned.} To the best of our knowledge, this unexpected behavior in $\mathcal{D}^{\text{folsom}}$ has not yet been discussed in the literature and, \redhighlight{upon further investigation, we found that the samples taken during time windows with fast moving cloud conditions (e.g.} \autoref{im: date_modif1}\redhighlight{) corresponded to the highest errors, leading us to discover a discrepancy between the date modified and the image file name timestamps.} In \autoref{im: date_modif1} we show \redhighlight{two distinct} 5 minute GHI/\redhighlight{DNI} windows in the middle row, with the image \textbf{date modified} alignment in the top row and the image \textbf{file name} alignment in the bottom row. \redhighlight{Both cases show} that the GHI began at a medium-to-low value, \redhighlight{which indicates some cloud coverage, before rapidly increasing, which indicates loss of cloud coverage and a more visible Sun. This behavior seems much more consistent with the top row images, strongly suggesting that it is actually the date modified timestamps that correctly correspond to the GHI time series. However, due to the strong presence of clouds, one could argue that this increase in GHI could be due to an extreme overirradiance event. However, since the DNI component is also being plotted, we can rule out this possibility and confirm that the Sun should indeed transition from covered to visible during both time windows.} 

\begin{figure}
\centering
\includegraphics[width=\linewidth]{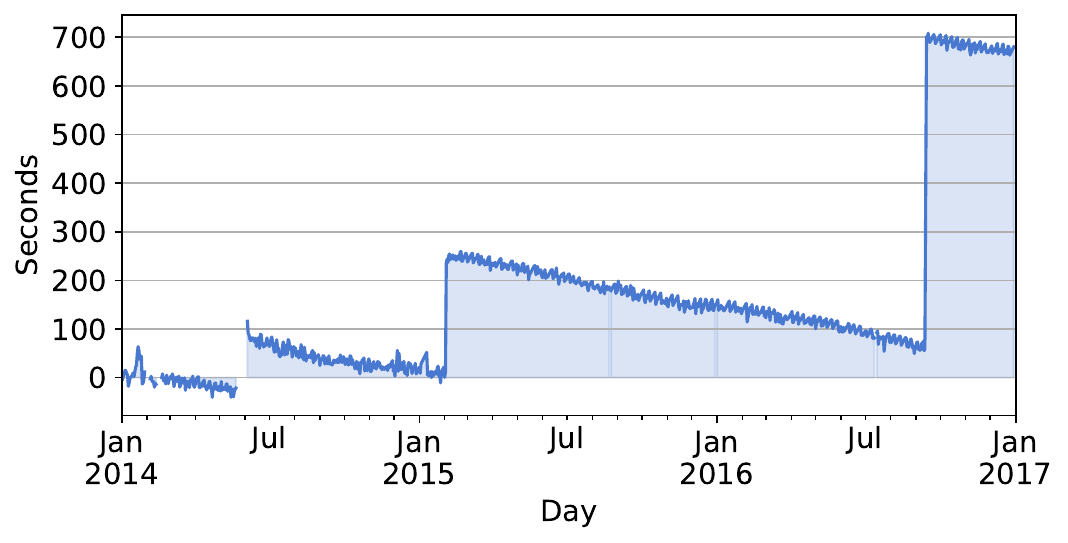}
\caption{Daily average difference between the file name and date modified of the image data in  $\mathcal{D}^{\text{folsom}}$. A positive value indicates that the file name is ahead of the date modified and vice-versa.}\label{im: date_modif2}
\end{figure}

In the 5-minute windows being shown in \autoref{im: date_modif1}, the file name is approximately \redhighlight{11 minutes 30 seconds} ahead of the date modified, but, from our analysis, we found that this difference \redhighlight{changes} throughout the dataset, as depicted in \autoref{im: date_modif2}, \redhighlight{which clearly indicates a time drift between both timestamps}\footnote{We hypothesize that the file name timestamps are based on the camera's internal clock, while the date modified is based on the database's clock. Assuming the irradiance timestamps are synchronized with the latter, then the file name timestamp would indeed result in misaligned image-irradiance pairs.}.
\redhighlight{Since the date modified seems much better aligned with the GHI time series, we use it to label the images in $\mathcal{D}^{\text{folsom}}$, except during the experiment in} \autoref{image_timestamps}\redhighlight{, where we provide a more quantitative analysis on this issue.}

As a precautionary measure, we also compared the metadata of $x^{\text{sirta}}$ and $x^{\text{nrel}}$ to their file names, \redhighlight{and found that the NREL images also contain an irregularity. However, the date modified issue in this dataset seems much more complex than in $\mathcal{D}^{\text{folsom}}$. In some cases, aligning the images with the date modified yielded a more sensible irradiance label, while in others, the opposite was true. This led us to choose the most simple solution possible, in which we discard all samples where} $|\text{Timestamp}(\text{Date Modified}) - \text{Timestamp}(\text{File Name})| > 30 \text{ s}$. \redhighlight{This processing step retains 65,202 of the original 66,908 image/irradiance pairs and ensures a more proper alignment for all retained pairs.} 

\begin{figure}
\centering
\includegraphics[width=\linewidth]{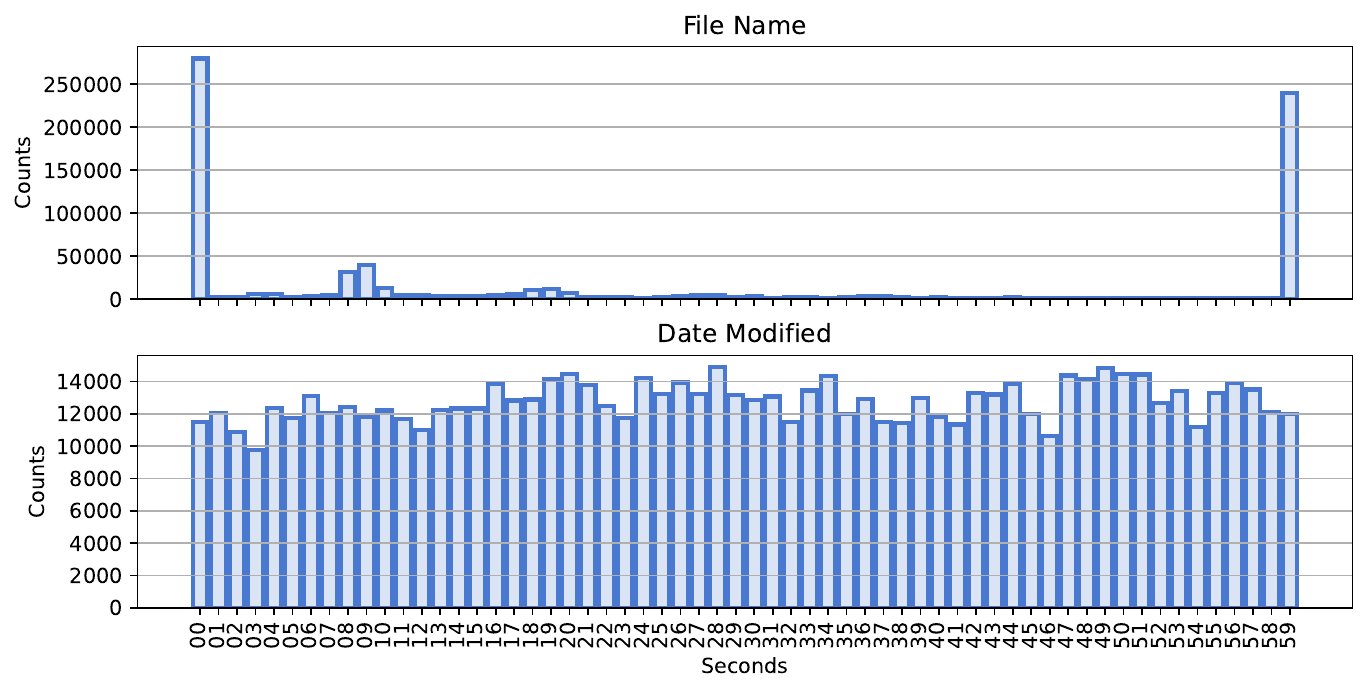}
\caption{Image timestamp second counts in $\mathcal{D}^{\text{folsom}}$. The upper axis shows that the file name timestamps are mostly synchronized at either :00 or :59 seconds (i.e. the file ends in ``00.jpg'' or ``59.jpg''). On the other hand, the lower axis shows that the date modified timestamps are uniformly spaced throughout the 60 possible values.}\label{im: sinc_img}
\end{figure}

Another thing to keep in mind when aligning the images with the irradiance data is the irregularity in the image timestamps, which is especially true assuming the date modified as the correct timestamp (see \autoref{im: sinc_img}). Since the irradiance samples are all synchronized at :00 seconds, aligning an image that was captured at :30 seconds with the nearest irradiance sample could result in a large error, especially in fast-moving cloud conditions. To mitigate this problem, we linearly interpolate the irradiance samples before the alignment. While this mostly affects $\mathcal{D}^{\text{folsom}}$, the linear interpolation is applied to the other two datasets as well.

\subsection{Sun Mask Generation}
\label{sun_mask_generation}
\begin{figure}
\centering
\includegraphics[width=\linewidth]{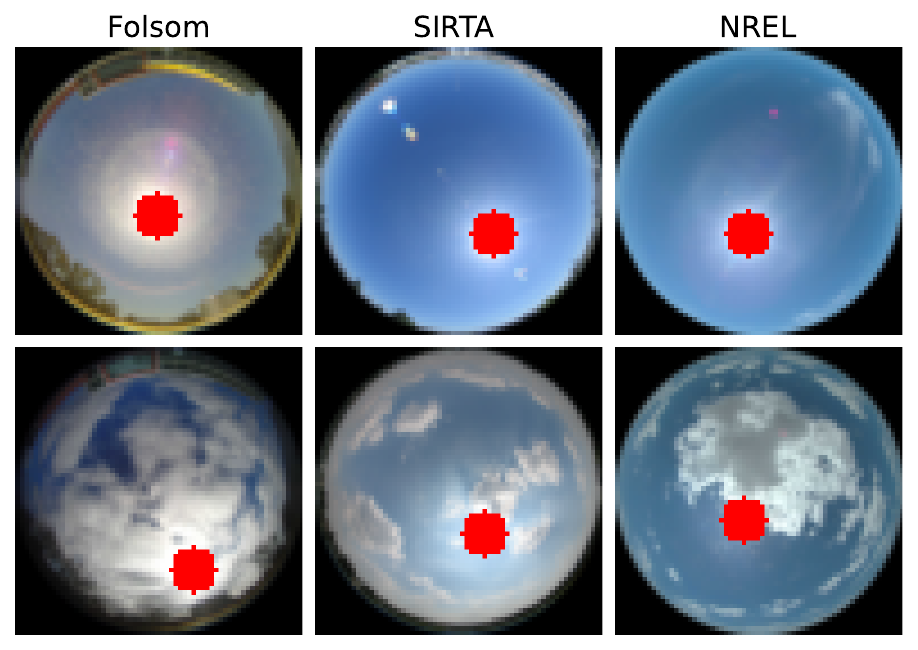}
\caption{Sun mask examples applied to the 64 $\times$ 64 images in all three datasets under cloudy and clear sky conditions. For a 64 $\times$ 64 image, the circle has a 5 pixel radius and is centered on $\text{x}_p$ and $\text{y}_p$.}\label{im: sun_mask_examples}
\end{figure}

In \autoref{sun_mask} we conduct an ablation experiment where we use a Sun mask as an additional input to the \redhighlight{estimation} model, which has shown to significantly benefit the performance in \cite{Papatheofanous.etal.2022.Deep-Learning-Based-Image}. First, we compute the distance of the Sun center from the center of the image plane, denoted by R, which is a function of the focal length of the camera \textit{f} and the angle from the optical axis $\Phi$. This mapping function will vary, since different types of lenses will have different projection types \cite{fisheye_lens}. For the stereographic projection, the mapping function is $\text{R}=2f\tan{\frac{\Phi}{2}}$, and for the equidistant projection the mapping function is $\text{R}=f\Phi$, where $\Phi$ is replaced by the zenith angle $\theta_z$. Following this, we compute the Sun center coordinates on the image plane ($\text{x}_c$, $\text{y}_c$) according to: 
\begin{equation}
\label{eq: x_center_img}
\text{x}_c = \text{R}\sin(\theta_a - \theta_c)
\end{equation}
\begin{equation}
\label{eq: y_center_img}
\text{y}_c = \text{R}\cos(\theta_a - \theta_c)
\end{equation}
where $\theta_a$ is the azimuth angle and $\theta_c$ is a correction angle that compensates for the orientation of the camera. To transform $\text{x}_c$ and $\text{y}_c$ to pixel coordinates and obtain $\text{x}_p$ and $\text{y}_p$, we apply:
\begin{equation}
\text{x}_p = \frac{W}{2}(1+\text{x}_c)
\end{equation}
\begin{equation}
\text{y}_p = \frac{W}{2}(1+\text{y}_c)
\end{equation}
where $W$ is the image width. Finally, we create the binary Sun mask as a circle with a 5 pixel radius (for $W=64$) and centered on $\text{x}_p$ and $\text{y}_p$, as shown in \autoref{im: sun_mask_examples}.

In the original paper, they only apply this mask in $\mathcal{D}^{\text{folsom}}$, in which they determine the stereographic projection for the R mapping function, $f=0.48$ and $\theta_c=165^{\circ}$. However, using these same parameters in $\mathcal{D}^{\text{sirta}}$ and $\mathcal{D}^{\text{nrel}}$ yielded bad results, so, through trial and error, we tuned them according to \autoref{tab:sun_mask_params}. The other two parameters, $\theta_z$ and $\theta_a$, are obtained with Python's pvlib library. For more details on this algorithm, we refer the reader to the original paper \cite{Papatheofanous.etal.2022.Deep-Learning-Based-Image}.

\begin{table}
  \centering
  \caption{Sun center localization parameters for each dataset. \textit{f} is the focal length and $\theta_c$ is the correction angle.}
  \label{tab:sun_mask_params}
      \begin{tabular}{cccc}  
      \toprule
      Dataset & Projection & \textit{f} &$\theta_c$ \\
      \toprule
      Folsom & Stereographic & 0.48 & 165$^{\circ}$\\
      \cline{1-4}
      SIRTA & Equidistant & 0.63 & 182$^{\circ}$\\
      \cline{1-4}
      NREL & Equidistant & 0.67 & 180$^{\circ}$\\
      \cline{1-4}
      \bottomrule
      \end{tabular}
\end{table}

\subsection{Train, Test and Validation Splits}
\label{train_test_val_splits}
All three datasets are first split into a train set, which is used to optimize the parameters of the models, and a testing set, which is used to evaluate said models. To account for all possible seasonal variations, we select one full year for testing, which was 2016, 2019 and 2023 for $\mathcal{D}^{\text{folsom}}$, $\mathcal{D}^{\text{sirta}}$ and $\mathcal{D}^{\text{nrel}}$, respectively. Moreover, the algorithm described in \autoref{clear_sky_detection} is used to further split the testing set into two additional sets: a clear sky and a cloudy sky set. 

The remaining years of each data set are used for the train set. In order to optimize the hyperparameters of the model, we split the full train set into a smaller train set and a validation set using $K$-Fold cross validation (CV). More specifically, we use stratified group $K$-Fold (with $K=5$), where each day is assigned to a different group (to avoid leaking samples from the same day in the training/validation sets) and each bin in the original train set (see \autoref{im: irradiance_data}) is assigned to a different class to maintain the original irradiance distribution in all 5 folds. After optimizing the hyperparameters of each model, we use the full train set to train a final model, which will be evaluated on the test set. \autoref{tab:samples_table} shows the number of samples \redhighlight{and year} of each set, after the application of all the pre-processing steps.

\begin{table}
\centering
\caption{Number of samples \redhighlight{and year} of each set after the application of all the pre-processing steps.}
\label{tab:samples_table}
\resizebox{\columnwidth}{!}{%
  \begin{tabular}{lcccccc}  
      \toprule
      & \multicolumn{2}{c}{Train} &\multicolumn{4}{c}{Test} \\
      \cmidrule(r){2-3}\cmidrule(r){4-7}
      Dataset    & All & Year & All & Clear & Cloudy & Year \\
      \midrule
      Folsom & 426,515 & 2014-2015 & 220,755 & 117,423 & 103,332 & 2016\\
      \midrule
      SIRTA & 297,774 & 2017-2018 & 103,926 & 18,447 & 85,479 & 2019\\
      \midrule
      NREL & 48,329 & 2020-2022 & 16,873 & 6,973 & 9,900 & 2023\\
      \midrule
      \bottomrule
      \end{tabular}
}
\end{table}

\section{Methods}
\label{methods}
\begin{figure*}
\centering
\includegraphics[width=\linewidth]{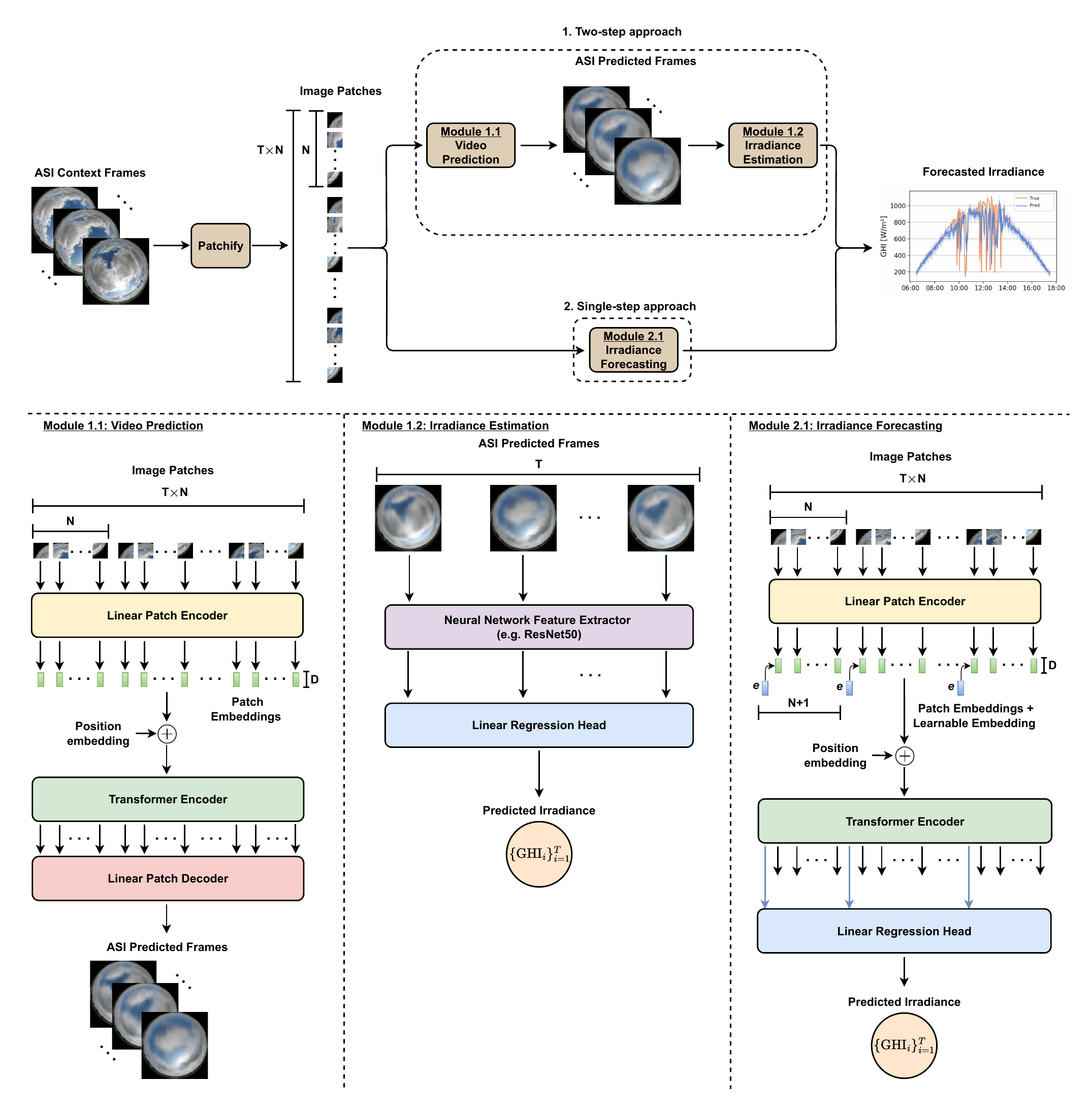}
\caption{\red{Two-step and single-step approaches for GHI nowcasting. Given an ASI sequence input of $T$ frames, each frame is transformed into a sequence of $N$ non-overlapping patches. In the two-step approach, these patch sequences will be fed into a video prediction module that predicts future ASI frames, and a subsequent module estimates the corresponding irradiance value for each generated frame. It is important to note that this irradiance estimation module is trained independently on the ground truth images. In the single-step approach, the irradiance forecasting module directly predicts the future irradiance values, by applying a linear regression head to the output of the learnable patch embedding, which we prepend to all $T$ patch sequences.}}\marginnote{\red{Rev.3 Com.3}} \label{im: two_step_approach}
\end{figure*}

\redhighlight{Our study \red{investigates} a two-step forecasting approach, akin to SkyGPT} \cite{Nie.etal.2024.SkyGPT-Probabilistic-Ultra-short-term}\redhighlight{, depicted in} \autoref{im: two_step_approach}\redhighlight{. A compelling aspect of this approach is that both modules can be designed and optimized entirely independently, i.e., the approach completely decouples future event prediction and GHI estimation. Furthermore, this approach allows us to obtain a lower bound on the GHI forecasting error, simply by replacing the predicted future images with the ground truth future images. \red{Our experiments target primarily the GHI estimation module, however we also include proof of concept experiments where we compare the two-step and single-step forecasting approaches depicted in \autoref{im: two_step_approach}.}}

\redhighlight{In} \autoref{ghi_estimation_section}\redhighlight{, we detail our training setup, the model architectures we trained and the evaluation metrics used in the GHI estimation experiments. All of the GHI estimation models are trained and evaluated separately for each dataset, however we also evaluate the aggregated performance across all three datasets.}

\redhighlight{In} \autoref{forecasting}\redhighlight{, we briefly describe our training objectives and implementations details for the forecasting experiments, which are only evaluated on $\mathcal{D}^{\text{sirta}}$.}

\subsection{GHI Estimation}\label{ghi_estimation_section}
\subsubsection{Training Setup}
\label{training_setup}




Given an input image $x_i$, each model produces an estimate~$\hat{I}_i$ of the concurrent GHI measurement $I_i$. The models are trained to minimize the Mean Squared Error (MSE) loss function given by:
\begin{equation}
\label{eq: mse}
L_{\text{MSE}} = \frac{1}{B} \sum_{i=1}^B (y_i - \hat{y}_i)^2
\end{equation}
where $B$ is the batch size, $y_i$ is the ground truth label of the $i^{th}$ sample and $\hat{y}_i$ is the model's output for the $i^{th}$ sample. 
Following previous work \cite{Yang.etal.2021.3D-CNN-Based-Sky-Image, Paletta.etal.2024.Improving-Cross-site-Generalisability}, we use the clear sky index as our default target variable, $y_i = k_{t,i}$, which implies that the GHI estimate is computed as $\hat{I}_i = I^\text{clr} \hat{k}_{t,i}$, where $\hat{k}_{t,i} = \hat{y}_i$. Other target variables and loss functions are explored in \autoref{model_target}.

Our models are trained with a learning rate that exponentially decays from its initial value $\text{lr}_0$ to $\frac{\text{lr}_0}{10}$ during the first 75\% of training according to:
\begin{equation}
\label{eq: lr_decay}
\text{lr}_i = \text{lr}_{i-1} \times e^{\frac{\ln{1/10}}{0.75 \times\text{n\_epochs}}}
\end{equation}
where n\_epochs is the total number of epochs and $0 < i < 0.75 \times\text{n\_epochs}$. The remaining 25\% of training is done with the learning rate fixed at $\frac{\text{lr}_0}{10}$ and we start averaging the weights of the model at the end of every epoch, as this has shown to lead to better generalization \cite{Izmailov.etal.2019.Averaging-Weights-Leads}. Each model is trained on a NVIDIA GeForce RTX 3090 GPU (24 GB memory) for a total of 16 epochs, except during hyperparemeter optimization, where they are only trained for 8 epochs to save on computational time. Details on the initial learning rate and other hyperparameters values are described in the following section.  

\subsubsection{Model Architectures and Hyperparameter Optimization}
We train and compare 10 different model architectures in this study, 7 of which are popular deep learning architectures used in many computer vision tasks and 3 that were designed/adapted specifically for the solar task. The first 7 include: ResNets (18, 34 and 50) \cite{He.etal.2015.Deep-Residual-Learning}; VGG16 \cite{Simonyan.Zisserman.2015.Very-Deep-Convolutional}; EfficientNetV2-S \cite{Tan.Le.2021.EfficientNetV2-Smaller-Models}; RegNetY-1.6GF \cite{Radosavovic.etal.2020.Designing-Network-Design}; MobileNetV2 \cite{Sandler.etal.2019.MobileNetV2-Inverted-Residuals}. We use the Pytorch implementation of these models and always initialize them with their pre-trained weights, except for the final layer, since we change it to only predict a single output, corresponding to the irradiance value. The other 3 models already output a single value, since they were designed for the solar task, however they are randomly initialized since we were unable to access their pre-trained weights. They are: SolarNet \cite{Feng.Zhang.2020.SolarNet-Sky-Image-based}; Sunset \cite{Nie.etal.2023.SKIPPD-SKy-Images, Sun.etal.2019.Short-term-Solar-Power-1}; and a modified version of the popular U-Net \cite{Ronneberger.etal.2015.U-Net-Convolutional-Networks} architecture, modified in \cite{Nie.etal.2024.SkyGPT-Probabilistic-Ultra-short-term}. 

We search 30 different hyperparameter combinations per model per dataset, totaling 900 different searches, which are made through Bayesian optimization \cite{Frazier.2018.Tutorial-Bayesian-Optimization} using the train and validation splits defined in \autoref{train_test_val_splits} and with the goal of minimizing the average validation loss across all 5 folds. For hyperparameter optimization, we reduce the number of training epochs to 8 and only use 10\% of the training data in each dataset. The hyperparameter search space is shown in \autoref{tab:hyper_search}.

\begin{table}
  \centering
  \caption{Hyperparameter search space for all the models and datasets. The PyTorch weight initialization is made for all the layers up to the last in the 7 models already implemented in PyTorch. The other 3 models are always randomly initialized with a fixed seed.}
  \label{tab:hyper_search}
      \begin{tabular}{ll}  
      \toprule
      Hyperparameter & Searched Values \\
      \toprule
      Initial Learning Rate ($\text{lr}_0$) & $[1, 2, 3, 4, 5, 6, 7, 8, 9, 10]\times 10^{-4}$\\
      Weight Initialization ($\theta_0$) & PyTorch, Random\\
      Weight Decay ($\lambda$) & Uniform(0, $10^{-3}$)\\
      Batch Size ($B$) & 64, 128\\
      Optimizer (Op.) & SGD, Adam \cite{Kingma.Ba.2017.Adam-Method-Stochastic}, AdamW \cite{Loshchilov.Hutter.2019.Decoupled-Weight-Decay}\\
      Dropout (p) & 0.0, 0.1, 0.2, 0.3, 0.4, 0.5, 0.6, 0.7\\
      \midrule
      \bottomrule
      \end{tabular}
\end{table}

\subsubsection{Evaluation Metrics}
A large variety of metrics have been proposed by the solar forecasting community in order to assess the quality of a model's predictions in the most insightful way possible \cite{Zhang.etal.2015.Suite-Metrics-Assessing, Frias-Paredes.etal.2016.Introducing-Temporal-Distortion, Vallance.etal.2017.Standardized-Procedure-Assess}. In this study, we evaluate the root mean squared error (RMSE), the mean absolute error (MAE) \redhighlight{and the normalized RMSE (nRMSE)}, which are well established metrics for evaluating the overall fit of the predictions:
\begin{equation}
\text{RMSE} = \sqrt{\frac{1}{N}\sum_{i=1}^N (I_i - \hat{I}_i)^2}
\end{equation}
\begin{equation}
\text{MAE} =\frac{1}{N}\sum_{i=1}^N |I_i - \hat{I}_i|
\end{equation}
\begin{equation}
\text{nRMSE} =\frac{\text{RMSE}}{\frac{1}{N}\sum_{i=1}^N I_i} 
\end{equation}
where $N$ is the number of samples, $I_i$ is the ground truth measurement of the GHI for the $i^{th}$ sample and $\hat{I}_i$ is the model's estimate of the GHI for the $i^{th}$ sample.
We employ the RMSE as our main ranking metric, since it penalizes larger errors more heavily than the MAE. \redhighlight{The nRMSE is used specifically to analyze performance across different weather seasons and hours of the day, due to the scale-invariant nature of this metric.} 

Note that, regardless of our choice for the model's target variable, the evaluation metrics are always expressed in terms of the GHI ($\text{W/m}^2$) \redhighlight{or as a percentage of GHI (in the case of nRMSE),} as that is the variable of interest.

\subsection{Forecasting}\label{forecasting}

\red{In the forecasting task, we use $\mathcal{D}^{\text{sirta}}$ with the goal of generating future GHI values with lead times ranging from 2 to 10 minutes, based on 8 minutes of past ASI context. \marginnote{\red{Rev.3 Com.2}}We investigate two approaches: a single-step approach, where an irradiance forecasting module directly predicts the future irradiance values, and a two-step approach, where a video prediction module first generates future ASI frames, and a subsequent module estimates the corresponding irradiance values for each generated frame. This irradiance estimation module corresponds to our best GHI estimation module, which is trained on the ground truth images.}


\subsubsection{Data Processing}

\red{In order to create the ASI sequence samples, we closely followed the data processing pipeline in \cite{Paletta.etal.2022.ECLIPSE-Envisioning-CLoud}, which is used as the baseline for our forecasting results. This lead to a few modifications to the data processing steps used in our GHI estimation experiments, described in \autoref{data_desc}.} Briefly, the key differences are: 1) the test set only includes samples that correspond to the odd days of 2019; 2) each image in the input sequence has a $128\times128$ resolution.

Regarding the use of only odd days of 2019 as the test set, this only means that the test set in the \red{GHI} forecasting \red{experiments} will be about half the size of the test set in the GHI estimation \red{experiments}. \red{We note, however, that the training set years remain the same, which means that, for the two-step approach, there is no data leakage between the video prediction and the GHI estimation modules.}

\red{Regarding the use of $128\times 128$ image resolution, this means that we need to resize the generated images to $64\times 64$ prior to the GHI estimation module in the two-step approach.}

\subsubsection{Two-step Approach}
Given an input batch of image sequences $\red{X_i} \in \mathbb{R}^{B\times T\times C\times H\times W}$, \redhighlight{the model is trained to predict the output image sequence} $\red{Y_i} \in \mathbb{R}^{B\times T\times C\times H\times W}$ \redhighlight{by minimizing the pixel-wise MSE, where $B$ is the batch size, $T$ is the number of frames, $C$ is the number of image channels, $H$ is the image height and $W$ is the image width. We use $T=5$ frames, such that the input consists of frames at times $\{t-8, t-6, t-4, t-2, t\}$, and the model predicts frames at $\{t+2, t+4, t+6, t+8, t+10\}$, in minutes. Our batch size is set to 16.}

\redhighlight{We employ an encoder-only Vision Transformer, closely following the design of PredFormer} \cite{tang2025predformer}, \redhighlight{which transforms the input image sequence $\red{X_i}$ into a sequence of $T\cdot N$ non-overlapping patches, where $N=\frac{HW}{p^2}$ is the number of patches per frame and $p$ is the patch size. We use a patch size of 16, resulting in $N=64$ patches per frame. Each image patch gets transformed into a patch embedding of dimension $D=256$, resulting in a $\red{X_p} \in \mathbb{R}^{B\times T\times N\times D}$ tensor.}

\redhighlight{A positional encoding is then added to $\red{X_p}$ to provide both the spatial and temporal information of each patch. Following this, each patch embedding will go through the stack of PredFormer layers, before finally being decoded back into the image patches by the linear patch decoder layer, which decodes $\red{X_p}$ \red{into $Y_i$}.}

\begin{figure}
\centering
\includegraphics[width=0.5\linewidth]{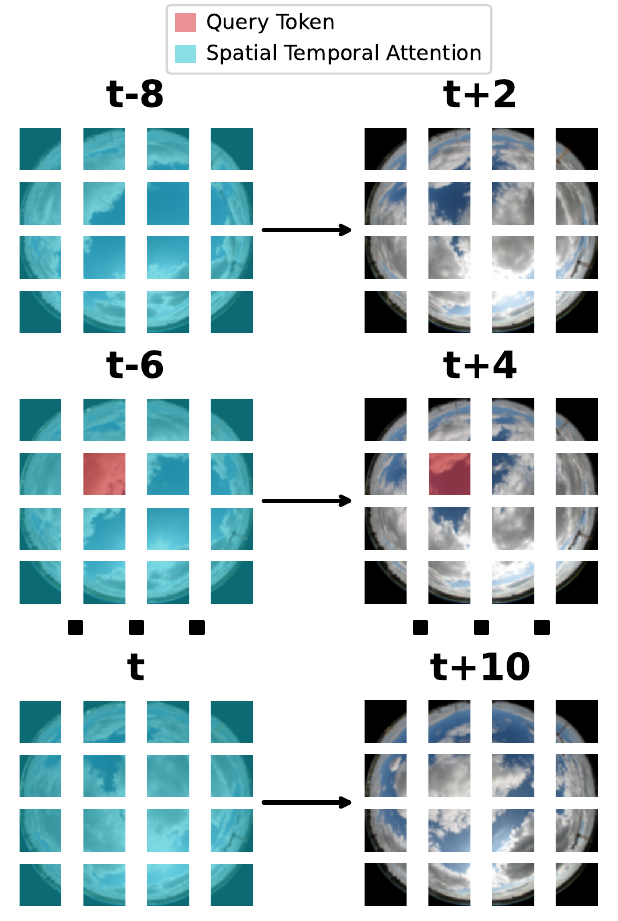}
\caption{\redhighlight{Input patches (left) and output patches (right) of the vision transformer model. Each query patch (red) can attend to all input patch tokens through full spatial temporal self-attention (blue). Each patch has size $16\times16$, however, for better visualization, we show patches of size $32\times32$ above.}} \label{im: transformer_details}
\end{figure}

\redhighlight{Regarding the PredFormer layers, we use full spatial-temporal self-attention, which allows each query patch to attend to all other patch tokens across the input sequence. Under this setup, each query patch will predict the patch at the exact same spatial position, but 10 minutes into the future, as illustrated by} \autoref{im: transformer_details}. \redhighlight{For more details on the model architecture, we refer the reader to the PredFormer article} \cite{tang2025predformer}.

\marginnote{\red{Rev.3 Com.1}}\red{Finally, to obtain the GHI predictions, each of the $T$ predicted frames in $\hat{Y}_i$ are resized to a $64\times 64$ resolution, and are then fed to our best pre-trained estimation model to obtain the corresponding GHI estimate.}

\subsubsection{Single-step Approach}

\marginnote{\red{Rev.3 Com.1}}\red{This approach is conceptually similar to the first step of the two-step approach, and also follows the design of Predformer. However, instead of training the model to predict the future image frames, it} is trained to directly predict the output clear sky index sequence $k_{t,i} \in \mathbb{R}^{B\times T}$ by minimizing the MSE
\begin{equation}
    L_{\text{MSE}} = \frac{1}{B\cdot T}\sum_{i=1}^B\sum_{t=1}^T\left(k_{t,i}-\hat{k}_{t,i}\right)^2
\end{equation}
\marginnote{\red{Rev.3 Com.1}}\redhighlight{where $B$ and $T$ have the exact same values used in the \red{first step of the} two-step approach. To optimize the PredFormer model with this loss function, two other changes \red{were made}. First, we prepend a learnable token embedding $e \in \mathbb{R}^{1\times 1\times 1\times D}$ to every frame of every batch sample in $\red{X_p}$, such that the new $\red{X_p} \in \mathbb{R}^{B\times T\times (N+1)\times D}$. This is analogous to the \textit{CLS} token used in BERT} \cite{DBLP:journals/corr/abs-1810-04805}\redhighlight{. Second, the linear patch decoder layer is replaced with a linear irradiance regression head. \marginnote{\red{Rev.3 Com.1}}Again, akin to BERT, this linear regression layer is only applied to the output of the $e$ token, obtaining the clear sky index prediction, which we transform into a GHI estimate through} \autoref{eq: clear_sky_index_eq}\redhighlight{. Everything else is the same as in the \red{first step of the} two-step approach.}

\subsubsection{Evaluation Metrics}
\redhighlight{We evaluate the results at every timestep} $\{t+2, t+4, t+6, t+8, t+10\}$ \redhighlight{using both the RMSE and the forecasting skill (FS). The forecasting skill is a metric that measures the overall improvement over the SPM, according to}:
\begin{equation}
    \text{FS} = \left(1-\frac{\text{RMSE}}{\text{RMSE}_{\text{SPM}}}\right)
\end{equation}
\redhighlight{where $\text{RMSE}_{\text{SMP}}$ is the error of the SPM predictions, given by} \autoref{eq: smart_persistence_model}\redhighlight{. A positive FS implies that a model is better than the SPM and a negative value suggests otherwise.}

\section{Experiments and Results}
\label{results}
\subsection{Benchmarking Results}\label{model_architectures}
\autoref{tab: benchmark_results_table} shows the RMSE and MAE for the test set for all ten models and all three datasets, as well as the aggregated performance across all datasets. \redhighlight{From these results, we observe that model depth and the usage of residual connections improves performance. For example, the ResNet50 and EfficientNet models, which are both deep residual models, consistently outperformed the others, especially under cloudy sky conditions. This is expected, since deeper networks are better equipped to extract complex spatial features, which are crucial under these sky conditions, due to the high variety of shapes, sizes and possible overirradiance effects that clouds can have.}

\redhighlight{However, models that are deep but lack residual connections, such as SolarNet and VGG16, showed significantly worse performance. This aligns with the findings of} \cite{He.etal.2015.Deep-Residual-Learning}\redhighlight{, who demonstrated that residual connections facilitate the information and gradient flow through the network, allowing for a much more efficient training of very deep models. The shallowest models in our benchmark, Sunset and UNet, yielded the worst performance. Their limited depth is likely a determinant factor in this poor performance, since it restricts their ability to extract the complex spatial features of clouds.}

\redhighlight{Finally, the results show that $\mathcal{D}^{\text{folsom}}$ achieved the best ``All'' performance, despite yielding worst results on both the ``Clear'' and ``Cloudy'' subsets (an instance of Simpson's paradox). The best ``All'' performance can be attributed to the fact that over 53\% of the samples in $\mathcal{D}^{\text{folsom}}$ are clear sky samples (as depicted in} \autoref{tab:samples_table}\redhighlight{), which will naturally bias the overall error toward lower values. On the other hand, the performance on the other two subsets was much worse, even compared to $\mathcal{D}^{\text{nrel}}$, which has far fewer training samples. We hypothesize that this behavior is due to the lens contamination we highlighted in} \autoref{folsom_dataset_descp} and \autoref{im: dirty_samples}\redhighlight{, since the model can be mistaking the dirt stains for false clouds, hurting the performance on both the sky conditions.}

\begin{table*}
\caption{Model architecture benchmark results for all three datasets. The best performing architecture is highlighted in green. The Aggr. dataset refers to the aggregated predictions/ground truth for all three datasets.}
\label{tab: benchmark_results_table}
\begin{tabular}{llllllll}
\toprule
 &  & \multicolumn{3}{c}{RMSE [W/m²]} & \multicolumn{3}{c}{MAE [W/m²]} \\
 \cmidrule(r){3-5} \cmidrule(r){6-8}
 & Sky cond. & All & Cloudy & Clear & All & Cloudy & Clear \\
Dataset &  &  &  &  &  &  &  \\
\midrule
\multirow[t]{10}{*}{Folsom} & EfficientNet &  37.39 ±  0.12 &  \cellcolor{green!25}52.60 ±  0.18 &  13.86 ±  0.51 &  20.87 ±  0.25 &  \cellcolor{green!25}32.75 ±  0.20 &  10.40 ±  0.32 \\
 & MobileNetV2 &  45.17 ±  2.04 &  62.22 ±  2.11 &  20.60 ±  3.06 &  27.94 ±  1.93 &  41.67 ±  1.87 &  15.85 ±  2.20 \\
 & RegNet &  37.68 ±  0.36 &  53.21 ±  0.37 &  13.31 ±  0.79 &  20.91 ±  0.47 &  33.21 ±  0.39 &  10.09 ±  0.58 \\
 & ResNet18 &  40.26 ±  0.81 &  56.31 ±  1.17 &  16.02 ±  0.57 &  22.92 ±  0.50 &  35.18 ±  0.84 &  12.13 ±  0.37 \\
 & ResNet34 &  45.91 ±  11.74 &  54.34 ±  0.52 &  31.65 ±  24.91 &  22.84 ±  1.84 &  33.37 ±  0.17 &  13.57 ±  3.41 \\
 & ResNet50 &  \cellcolor{green!25}37.21 ±  0.26 &  52.62 ±  0.35 &  \cellcolor{green!25}12.89 ±  0.39 &  \cellcolor{green!25}20.58 ±  0.15 &  32.82 ±  0.25 &  \cellcolor{green!25}9.80 ±  0.23 \\
 & SolarNet &  43.25 ±  1.81 &  59.87 ±  1.70 &  18.90 ±  2.93 &  25.32 ±  1.43 &  38.12 ±  1.13 &  14.06 ±  1.69 \\
 & Sunset &  48.08 ±  2.31 &  65.34 ±  2.64 &  24.06 ±  3.70 &  29.16 ±  2.18 &  41.89 ±  1.92 &  17.96 ±  2.85 \\
 & UNet &  50.12 ±  2.27 &  64.29 ±  1.45 &  32.67 ±  5.14 &  33.57 ±  2.89 &  43.10 ±  1.47 &  25.18 ±  4.38 \\
 & VGG16 &  39.64 ±  0.52 &  55.35 ±  0.57 &  16.03 ±  0.78 &  22.91 ±  0.47 &  35.28 ±  0.46 &  12.02 ±  0.54 \\
\cline{1-8}
\multirow[t]{10}{*}{SIRTA} & EfficientNet &  40.19 ±  0.23 &  44.00 ±  0.27 &  11.42 ±  0.42 &  23.72 ±  0.21 &  27.00 ±  0.31 &  8.53 ±  0.43 \\
 & MobileNetV2 &  40.61 ±  0.27 &  44.44 ±  0.32 &  11.91 ±  1.00 &  23.56 ±  0.21 &  26.72 ±  0.22 &  8.94 ±  0.67 \\
 & RegNet &  45.25 ±  2.09 &  49.03 ±  1.88 &  19.29 ±  5.48 &  27.99 ±  1.70 &  31.05 ±  1.27 &  13.78 ±  3.74 \\
 & ResNet18 &  44.80 ±  2.24 &  48.66 ±  2.41 &  18.21 ±  2.08 &  27.90 ±  1.74 &  31.14 ±  1.94 &  12.89 ±  1.22 \\
 & ResNet34 &  43.72 ±  2.00 &  46.60 ±  1.22 &  25.32 ±  9.00 &  26.71 ±  1.47 &  28.95 ±  0.91 &  16.31 ±  4.28 \\
 & ResNet50 &  \cellcolor{green!25}39.68 ±  0.21 &  \cellcolor{green!25}43.46 ±  0.22 &  \cellcolor{green!25}10.75 ±  0.56 &  \cellcolor{green!25}23.06 ±  0.16 &  \cellcolor{green!25}26.30 ±  0.14 &  \cellcolor{green!25}8.07 ±  0.42 \\
 & SolarNet &  42.99 ±  0.58 &  47.04 ±  0.63 &  12.57 ±  0.44 &  25.21 ±  0.43 &  28.64 ±  0.47 &  9.31 ±  0.37 \\
 & Sunset &  46.90 ±  1.03 &  50.97 ±  0.94 &  18.77 ±  2.81 &  29.46 ±  0.96 &  32.81 ±  0.80 &  13.94 ±  1.96 \\
 & UNet &  47.75 ±  1.44 &  51.31 ±  1.50 &  25.36 ±  1.72 &  30.68 ±  0.96 &  33.33 ±  0.92 &  18.43 ±  1.29 \\
 & VGG16 &  52.57 ±  2.81 &  56.16 ±  2.50 &  30.62 ±  5.96 &  33.67 ±  2.30 &  36.30 ±  1.74 &  21.48 ±  4.92 \\
\cline{1-8}
\multirow[t]{10}{*}{\redhighlight{NREL}} & EfficientNet &  42.44 ±  4.58 &  52.14 ±  1.64 &  19.21 ±  13.28 &  23.97 ±  1.25 &  33.43 ±  1.11 &  10.53 ±  1.57 \\
 & MobileNetV2 &  47.18 ±  1.18 &  60.57 ±  1.49 &  13.31 ±  0.53 &  27.15 ±  0.58 &  39.36 ±  0.77 &  9.80 ±  0.33 \\
 & RegNet &  41.65 ±  0.33 &  53.01 ±  0.44 &  14.44 ±  0.49 &  24.42 ±  0.17 &  33.47 ±  0.19 &  11.58 ±  0.29 \\
 & ResNet18 &  40.84 ±  0.57 &  52.36 ±  0.75 &  11.91 ±  1.76 &  23.17 ±  0.54 &  33.12 ±  0.37 &  9.03 ±  1.25 \\
 & ResNet34 &  43.12 ±  3.10 &  53.56 ±  3.19 &  20.25 ±  4.87 &  26.05 ±  2.57 &  34.00 ±  2.42 &  14.77 ±  3.11 \\
 & ResNet50 &  \cellcolor{green!25}38.75 ±  0.29 &  \cellcolor{green!25}49.85 ±  0.37 &  \cellcolor{green!25}10.25 ±  0.58 &  \cellcolor{green!25}21.73 ±  0.23 &  \cellcolor{green!25}31.56 ±  0.38 &  \cellcolor{green!25}7.77 ±  0.41 \\
 & SolarNet &  44.33 ±  0.83 &  56.84 ±  0.98 &  12.93 ±  1.07 &  25.58 ±  0.84 &  36.62 ±  0.79 &  9.92 ±  1.08 \\
 & Sunset &  55.47 ±  2.76 &  71.24 ±  3.46 &  15.46 ±  1.42 &  31.14 ±  1.30 &  45.40 ±  1.79 &  10.89 ±  0.66 \\
 & UNet &  52.21 ±  1.77 &  66.45 ±  2.30 &  18.05 ±  1.07 &  31.43 ±  0.96 &  44.35 ±  1.60 &  13.09 ±  0.87 \\
 & VGG16 &  41.45 ±  0.95 &  53.19 ±  1.19 &  11.80 ±  0.81 &  23.62 ±  0.69 &  33.76 ±  0.73 &  9.24 ±  0.75 \\
\cline{1-8}
\multirow[t]{10}{*}{\redhighlight{Aggregated}} & EfficientNet &  38.53 ±  0.31 &  49.07 ±  0.12 &  14.13 ±  1.48 &  21.89 ±  0.22 &  30.31 ±  0.19 &  10.17 ±  0.30 \\
 & MobileNetV2 &  43.95 ±  1.34 &  55.19 ±  1.24 &  19.39 ±  2.71 &  26.57 ±  1.25 &  35.13 ±  0.99 &  14.66 ±  1.86 \\
 & RegNet &  40.33 ±  0.86 &  51.45 ±  0.90 &  14.37 ±  1.39 &  23.24 ±  0.73 &  32.29 ±  0.70 &  10.64 ±  0.83 \\
 & ResNet18 &  41.74 ±  0.67 &  52.98 ±  0.88 &  16.17 ±  0.26 &  24.45 ±  0.46 &  33.34 ±  0.68 &  12.08 ±  0.22 \\
 & ResNet34 &  45.41 ±  8.03 &  51.12 ±  0.46 &  31.40 ±  21.50 &  24.17 ±  1.30 &  31.50 ±  0.29 &  13.98 ±  3.00 \\
 & ResNet50 &  \cellcolor{green!25}38.05 ±  0.20 &  \cellcolor{green!25}48.75 ±  0.24 &  \cellcolor{green!25}12.51 ±  0.36 &  \cellcolor{green!25}21.39 ±  0.14 &  \cellcolor{green!25}29.95 ±  0.17 &  \cellcolor{green!25}9.48 ±  0.22 \\
 & SolarNet &  43.23 ±  1.36 &  54.56 ±  1.21 &  17.96 ±  2.59 &  25.30 ±  1.06 &  33.97 ±  0.79 &  13.25 ±  1.45 \\
 & Sunset &  48.14 ±  1.35 &  59.94 ±  1.44 &  23.15 ±  3.06 &  29.35 ±  1.23 &  38.16 ±  0.86 &  17.10 ±  2.27 \\
 & UNet &  49.54 ±  1.35 &  59.18 ±  0.97 &  31.29 ±  4.41 &  32.58 ±  1.78 &  38.96 ±  0.79 &  23.71 ±  3.60 \\
 & VGG16 &  44.08 ±  1.06 &  55.61 ±  1.11 &  18.50 ±  1.42 &  26.22 ±  0.78 &  35.64 ±  0.79 &  13.11 ±  0.79 \\
\cline{1-8}
\bottomrule
\end{tabular}
\end{table*}

\begin{figure}
\centering
\includegraphics[width=\linewidth]{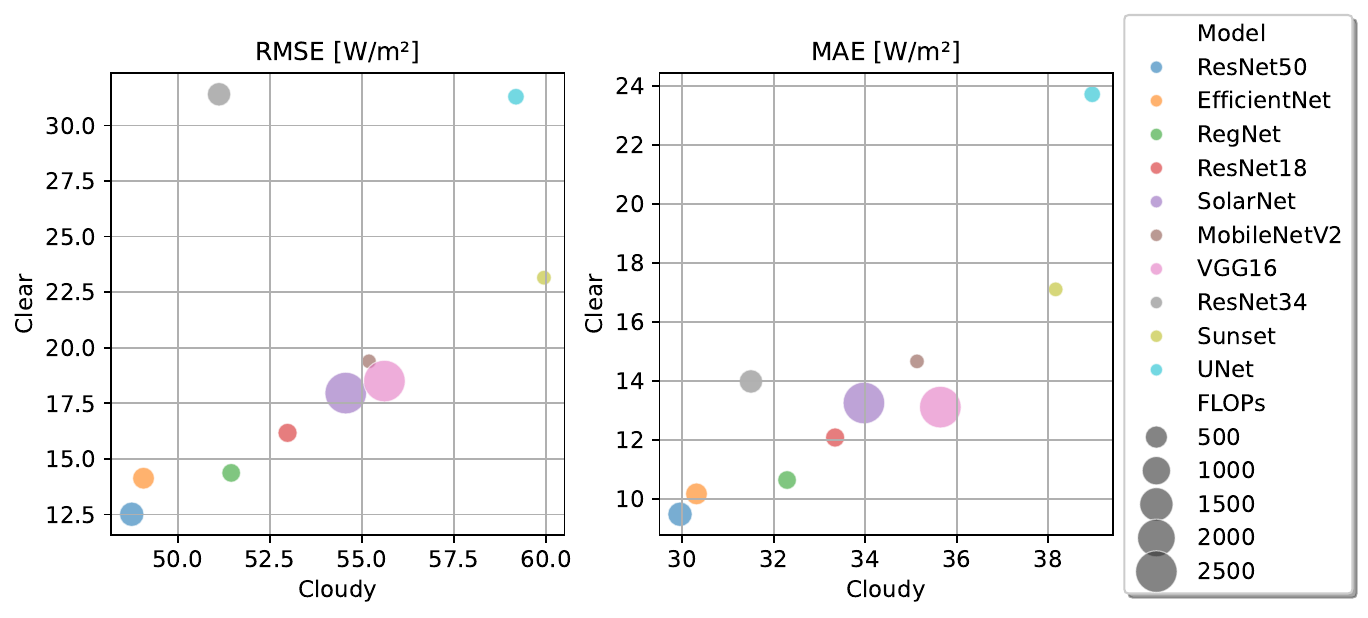}
\caption{Cloudy and clear sky scatter plot showing the aggregated performance of all ten models. The left panel shows the RMSE performance and the right panel shows the MAE. Marker sizes are scaled according to the number of FLOPs for each model.} \label{im: models_scatter_plot}
\end{figure}

\autoref{im: models_scatter_plot} summarizes the aggregated results in \autoref{tab: benchmark_results_table} for both the clear and cloudy sky test subsets.

\subsection{Ablation Experiments}
Based on the results from the previous section, we then perform several ablation experiments using the ResNet50 model with the goal of investigating the impact that different configurations of pre-processing steps and training setups have on the performance. These configurations are summarized in \autoref{tab: configs_table}. Note that, for the ``$\Delta \text{t}$'' and ``Sample Interval'' configuration, the searched values \textbf{are only applied to the training set}, while the default values are used for the test set.

\begin{table*}[t]
  \centering
  \caption{Configuration table for all the ablation experiments in this study. The values for a certain configuration are only searched in their respective sections, otherwise the default value is used.}
  \label{tab: configs_table}
      \begin{tabular}{lccc}  
      \toprule
      Configuration & Dataset & Searched Values & Default Values \\
      \toprule
       & Folsom & File Name, Date Modified & Date Modified\\
      \textbf{Image Timestamps} (\ref{image_timestamps}) & SIRTA & - & File Name\\
       & NREL & - & File Name\\
      \midrule
       & Folsom & -30s:10s:30s & -20s\\
      \textbf{Time off-set $\Delta \text{t}$} (\ref{timedelta}) & SIRTA & -30s:10s:30s & 10s\\
      & NREL & -30s:10s:30s & \redhighlight{-30s}\\
      \midrule
      & Folsom & & \\
      \textbf{Model Target} (\ref{model_target})& SIRTA & GHI, $k_t$, $K_t$, $k_t^{\text{weighted}}$, $K_t^{\text{weighted}}$& $k_t$\\
      & NREL & &\\
      \midrule
      & Folsom & Long & \\
      \textbf{Camera Exposure} (\ref{camera_exposure})& SIRTA & Short, Long & Long\\
      & NREL & Short, Long &\\
      \midrule
      & Folsom & 1 - 10 min & 1 min\\
      \textbf{Sample Interval} (\ref{sample_interval})& SIRTA & 1/2 - 10 min & 1/2 min\\
      & NREL & 10 min & 10 min\\
      \midrule
      & Folsom & &\\
      \textbf{Sun Mask} (\ref{sun_mask})& SIRTA & Mask, No Mask& No Mask\\
      & NREL & & \\
      \midrule
      \bottomrule
      \end{tabular}
\end{table*}

\subsubsection{Image Timestamps in $\mathcal{D}^{\text{folsom}}$}
\label{image_timestamps}

\begin{table}
  \centering
  \caption{\textbf{GHI} and \redhighlight{\textbf{DNI}} RMSE results for the four possible train/test combinations (FN = File Name; DM = Date Modified) and the different test subsets in $\mathcal{D}^{\text{folsom}}$.}
  \label{tab:img_timestamp_results}
  \begin{tabular}{lcccc}  
  \toprule
  & & & \multicolumn{2}{c}{RMSE [$\text{W/m}^2$]} \\
  \cmidrule(r){4-5}
  Sky Cond. & Train & Test & GHI & \redhighlight{DNI}\\
  \toprule
  All & FN & FN & 62.52 ± 0.18 & 96.04 ± 0.75\\
      & FN & DM & 56.72 ± 0.13 & 81.31 ± 1.28\\
      & DM & FN & 59.37 ± 0.34 & 96.60 ± 0.45\\
      & DM & DM & \textbf{37.21 ± 0.26} & \textbf{55.42 ± 0.94}\\
  \cline{1-5}
  Cloudy & FN & FN & 89.76 ± 0.35 & 135.81 ± 0.94\\
         & FN & DM & 81.19 ± 0.18 & 113.88 ± 1.66\\
         & DM & FN & 85.34 ± 0.48 & 135.39 ± 0.36\\
         & DM & DM & \textbf{52.62 ± 0.35} & \textbf{73.25 ± 0.86}\\
  \cline{1-5}
  Clear & FN & FN & 16.05 ± 0.60 & 33.13 ± 3.50\\
        & FN & DM & 15.76 ± 0.57 & \textbf{31.77 ± 3.49}\\
        & DM & FN & 14.69 ± 0.43 & 37.55 ± 1.60\\
        & DM & DM & \textbf{12.89 ± 0.39} & 32.42 ± 1.67\\
  \cline{1-5}
  \bottomrule
  \end{tabular}
\end{table}


As mentioned in \autoref{image_alignment}, aligning the sky images with the irradiance data in $\mathcal{D}^{\text{folsom}}$ using the image file names led to large label errors, \redhighlight{while the date modified seemed much more aligned with the irradiance data}. To the best of our knowledge, this has not yet been reported in the literature so, in order to ascertain that the date modified is truly the correct timestamp for labeling the images in $\mathcal{D}^{\text{folsom}}$, a more quantitative analysis is also needed. To this end, we train two different models, one trained on a file name aligned training set and another on a date modified aligned training set. Both of these models are then evaluated on file named aligned testing set and a date modified aligned testing set, totaling four different train/test image timestamp alignment combinations. All other configurations are kept at their default values, depicted in \autoref{tab: configs_table}, \redhighlight{and, for this particular experiment, we also evaluate the results on the DNI component, both to rule out the overirradiance hypothesis raised in} \autoref{image_alignment}\redhighlight{, as well as to offer a benchmark for future studies that use $\mathcal{D}^{\text{folsom}}$ to focus on this specific component.} 

\autoref{tab:img_timestamp_results} \redhighlight{depicts the GHI and DNI results. One can see that, when the training set and the testing set are aligned with the date modified timestamp, the RMSE drops significantly, confirming the assumptions made in} \autoref{image_alignment}. Perhaps unsurprisingly, this drop was much larger for the cloudy subset than the clear subset, since, for the majority of clear sky samples, a few minutes difference in alignment results in small GHI\redhighlight{/DNI} differences. On the other hand, under cloudy sky conditions, the sky image and irradiance alignment is much more critical.

Moreover, \autoref{tab:img_timestamp_results} also highlights the fact that evaluating with the date modified alignment is always better than with the file name alignment, even when training with the file name. The reason for this can be attributed to \autoref{im: date_modif2}, which shows that the file name alignment is much worse in 2016 than in 2014/2015. Since we selected 2014/2015 for training and 2016 for testing, this means that the model can still learn somewhat useful features when trained with the file name alignment.

\subsubsection{Time Shifting Operation}
\label{timedelta}
As described in \autoref{irr_data_proc}, we shift the irradiance measurements of the training set by applying a timestamp offset of $\Delta \text{t}$. For each dataset, we optimize $\Delta \text{t} \in \{-30\text{s} : 10\text{s} : 30\text{s} \}$ by applying a 5-fold CV (as described in \autoref{training_setup}) and minimizing the validation loss. We then apply the optimal $\Delta \text{t}$ to the full training set to train the model and evaluate the results on the testing set.

\begin{figure}
\centering
\includegraphics[width=\linewidth]{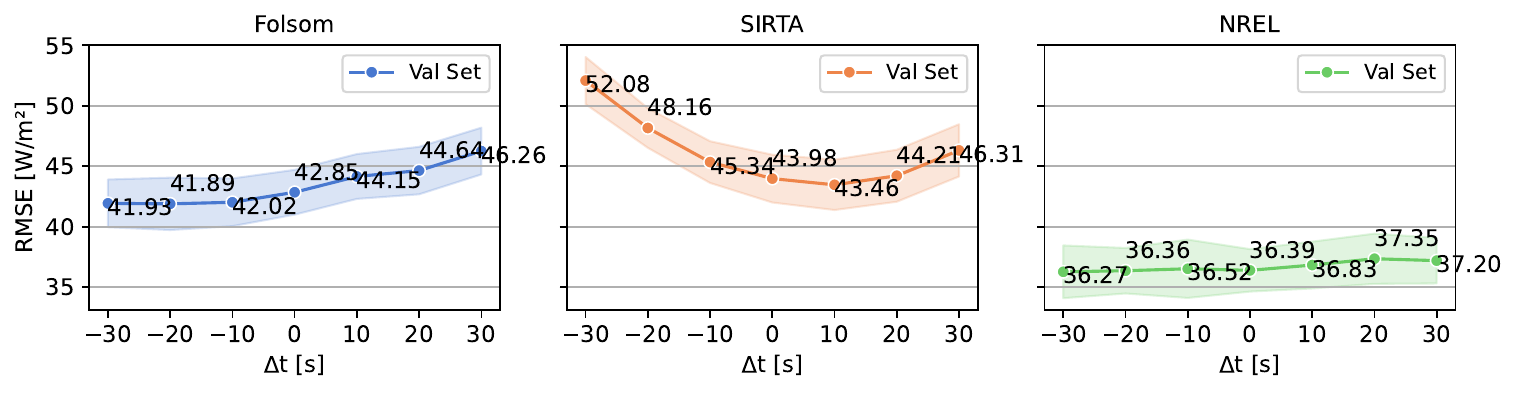}
\caption{\redhighlight{Validation set results for $\mathcal{D}^{\text{folsom}}$, $\mathcal{D}^{\text{sirta}}$ and $\mathcal{D}^{\text{nrel}}$. For each $\Delta \text{t}$, we report the mean RMSE across all 5 folds, as well as the standard deviation.}} \label{im: cv_results_timedelta}
\end{figure}

The CV results are depicted in \autoref{im: cv_results_timedelta}, which shows the mean validation set RMSE across all 5 folds, for each dataset. Interestingly, in $\mathcal{D}^{\text{folsom}}$ and $\mathcal{D}^{\text{nrel}}$, the validation RMSE decreases slightly for a negative $\Delta \text{t}$ and increases for a positive $\Delta \text{t}$, but in $\mathcal{D}^{\text{sirta}}$ the opposite happens. This suggests that the type of average applied to the GHI samples in $\mathcal{D}^{\text{folsom}}$ was the same as in $\mathcal{D}^{\text{nrel}}$, but different in $\mathcal{D}^{\text{sirta}}$. The optimal $\Delta \text{t}$ was -20s, 10s and \redhighlight{-30s} for $\mathcal{D}^{\text{folsom}}$, $\mathcal{D}^{\text{sirta}}$ and $\mathcal{D}^{\text{nrel}}$, respectively.

Regarding the results of the test set, \autoref{tab:timedelta_results} shows that using the optimal $\Delta \text{t}$ to train the models always leads to a consistent gain in performance for all datasets, \redhighlight{though in $\mathcal{D}^{\text{nrel}}$ this gain was close to negligible. The improvement is more apparent under cloudy sky conditions, since the sky image and irradiance alignment is more critical during these periods.}

\begin{table}
  \centering
  \caption{Test set performance comparison between a 0 second $\Delta \text{t}$ (baseline) and the best performing one, obtained during CV.}
  \label{tab:timedelta_results}
      \begin{tabular}{lcccc}  
      \toprule
      &  & \multicolumn{3}{c}{RMSE [W/m²]} \\ 
      \cmidrule(r){3-5}
      Dataset & $\Delta \text{t}$ & All & Clear & Cloudy \\
      \midrule
      Folsom & 0s & 40.24 ± 0.94 & 15.60 ± 1.36 & 56.41 ± 1.06 \\ 
             & -20s & \textbf{37.21 ± 0.26} & \textbf{12.89 ± 0.39} & \textbf{52.62 ± 0.35}\\
      \cline{1-5}
      SIRTA & 0s & 41.22 ± 0.54 & 11.17 ± 0.44 & 45.15 ± 0.59 \\
            & 10s & \textbf{39.68 ± 0.21} & \textbf{10.75 ± 0.56} & \textbf{43.46 ± 0.22}\\
      \cline{1-5}
      \redhighlight{NREL} & 0s & 39.01 ± 0.37  & 10.28 ± 0.53 & 50.20 ± 0.50\\
           & -30s& \textbf{38.75 ± 0.29} & \textbf{10.25 ± 0.58} & \textbf{49.85 ± 0.37}\\
     \cline{1-5}
      \bottomrule
      \end{tabular}
\end{table}

\subsubsection{Target Variable and Loss Function}
\label{model_target}
\begin{table*}[b]
\centering
\caption{Test set results for all 5 different model target variables. The best performing target is highlighted in green for each dataset.}
\label{tab: model_target_results_table}
\begin{tabular}{llllllll}
\toprule
 &  & \multicolumn{3}{c}{RMSE [W/m²]} & \multicolumn{3}{c}{MAE [W/m²]} \\
 \cmidrule(r){3-5}\cmidrule(r){6-8}
 & Sky cond. & All & Cloudy & Clear & All & Cloudy & Clear \\
Dataset &  &  &  &  &  &  &  \\
\midrule
\multirow[t]{5}{*}{Folsom} & GHI & 65.77 ± 2.22 & 82.62 ± 2.81 & 46.09 ± 1.80 & 48.24 ± 2.22 & 57.44 ± 2.29 & 40.14 ± 2.38 \\
 & $k_t$ & \cellcolor{green!25}37.21 ± 0.26 & \cellcolor{green!25}52.62 ± 0.35 & \cellcolor{green!25}12.89 ± 0.39 & \cellcolor{green!25}20.58 ± 0.15 & \cellcolor{green!25}32.82 ± 0.25 & \cellcolor{green!25}9.80 ± 0.23 \\
 & $K_t$ & 38.16 ± 0.62 & 53.09 ± 0.76 & 16.02 ± 0.90 & 22.09 ± 0.60 & 33.42 ± 0.70 & 12.12 ± 0.63 \\
 & $k_t$ (weighted) & 37.46 ± 0.15 & 52.87 ± 0.21 & 13.35 ± 0.45 & 21.02 ± 0.20 & 33.33 ± 0.32 & 10.19 ± 0.36 \\
 & $K_t$ (weighted) & 38.93 ± 0.66 & 53.87 ± 0.74 & 17.17 ± 0.89 & 23.06 ± 0.69 & 34.39 ± 0.71 & 13.08 ± 0.75 \\
\cline{1-8}
\multirow[t]{5}{*}{SIRTA} & GHI & 50.84 ± 2.18 & 54.29 ± 2.24 & 29.95 ± 3.10 & 34.69 ± 1.92 & 36.93 ± 1.91 & 24.30 ± 2.80 \\
 & $k_t$ & 39.68 ± 0.21 & 43.46 ± 0.22 & 10.75 ± 0.56 & 23.06 ± 0.16 & 26.30 ± 0.14 & 8.07 ± 0.42 \\
 & $K_t$ & 39.33 ± 0.43 & \cellcolor{green!25}42.91 ± 0.42 & 13.39 ± 1.14 & 23.13 ± 0.45 & \cellcolor{green!25}25.89 ± 0.35 & 10.33 ± 0.99 \\
 & $k_t$ (weighted) & 39.88 ± 0.31 & 43.72 ± 0.35 & \cellcolor{green!25}10.22 ± 0.42 & 23.34 ± 0.24 & 26.70 ± 0.31 & \cellcolor{green!25}7.78 ± 0.27 \\
 & $K_t$ (weighted) & \cellcolor{green!25}39.23 ± 0.28 & 42.94 ± 0.30 & 11.30 ± 0.80 & \cellcolor{green!25}22.97 ± 0.28 & 26.04 ± 0.26 & 8.71 ± 0.73 \\
\cline{1-8}
\multirow[t]{5}{*}{\redhighlight{NREL}} & GHI & 42.43 ± 0.68 & 54.12 ± 0.90 & 14.04 ± 0.37 & 25.94 ± 0.36 & 36.64 ± 0.58 & 10.74 ± 0.20 \\
 & $k_t$ & 38.75 ± 0.29 & 49.85 ± 0.37 & 10.25 ± 0.58 & 21.73 ±  0.23 & 31.56 ±  0.38 & 7.77 ±  0.41 \\
 & $K_t$ & 38.20 ± 0.24 & 48.91 ± 0.36 & 11.60 ± 0.48 & 21.97 ± 0.11 & 31.16 ± 0.19 & 8.91 ± 0.40 \\
 & $k_t$ (weighted) & 38.06 ± 0.33 & 48.94 ± 0.43 & \cellcolor{green!25}10.18 ± 0.51 & \cellcolor{green!25}21.45 ± 0.30 & \cellcolor{green!25}31.11 ± 0.41 & \cellcolor{green!25}7.72 ± 0.25 \\
 & $K_t$ (weighted) & \cellcolor{green!25}38.04 ± 0.26 & \cellcolor{green!25}48.78 ± 0.35 & 11.13 ± 0.18 & 21.83 ± 0.28 & 31.15 ± 0.36 & 8.60 ± 0.19 \\
\cline{1-8}
\bottomrule
\end{tabular}
\end{table*}
In this experiment, we consider five combinations of target variables and loss functions. Specifically, for the MSE loss \eqref{eq: mse}, we consider three choices for the target variable $y \in \{I, k_t, K_t\}$, resulting in the following expressions for the GHI estimate:
\begin{equation}
\label{GHI_estimate}
\hat{I}_i = \begin{cases}
\hat{y}_i, &\text{if $y_i = I_i$}, \\
I^\text{clr} \hat{k}_{t,i}, \; \text{where $\hat{k}_{t,i} = \hat{y}_i$}, &\text{if $y_i = k_{t,i}$}, \\
I^\text{extr} \hat{K}_{t,i}, \; \text{where $\hat{K}_{t,i} = \hat{y}_i$}, &\text{if $y_i = K_{t,i}$}.
\end{cases}
\end{equation}

\begin{figure}
\centering
\includegraphics[width=\linewidth]{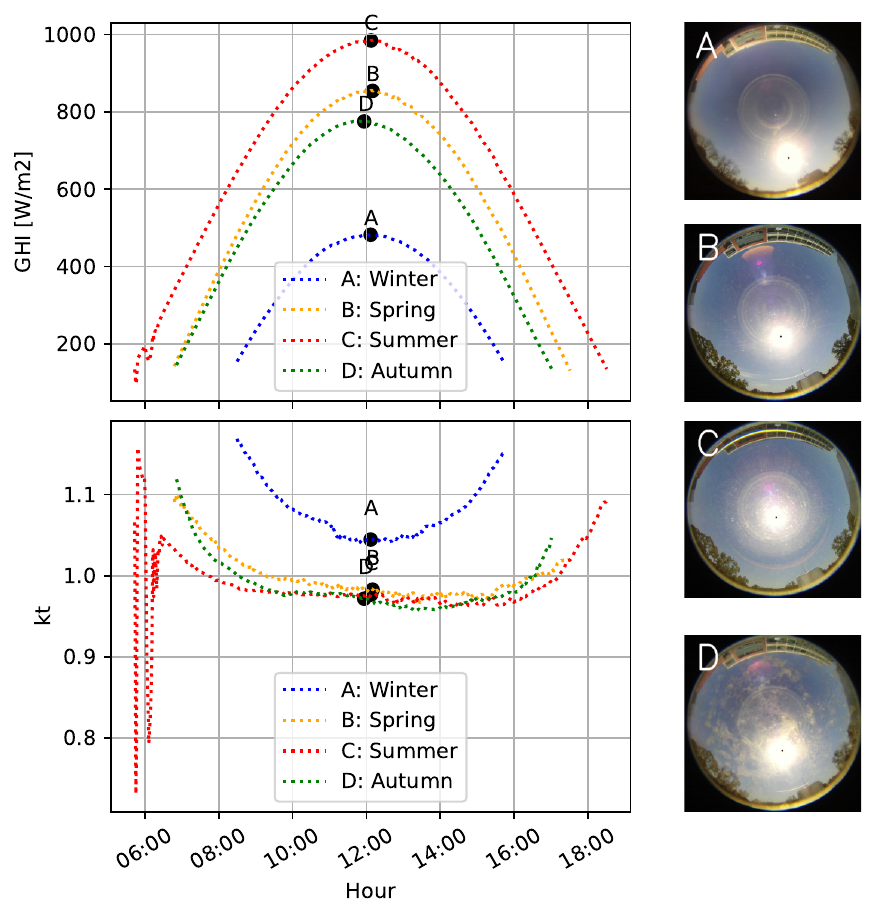}
\vspace{0.1ex}
\caption{GHI and $k_t$ values comparison for a clear sky day showing the seasonal variance in $\mathcal{D}^{\text{folsom}}$. The \red{upper} panel shows the GHI values and the \red{lower} panel shows the $k_t$ values. The corresponding images to each labeled marker \red{are displayed on the right.}} \label{im: model_target_example}
\end{figure}

Although using the raw GHI value as target is the most natural and simple choice, it suffers from the fact that the GHI values will have different scales across different weather seasons. The \red{upper} panel of \autoref{im: model_target_example} depicts a $\Delta$GHI of almost 500 W/m² between the peak GHI value in summer and in winter, which, from their corresponding images, can be inferred based on the distance from the Sun center to the image center. This means that models trained to predict raw GHI values need to encode the Sun's position in the image with extreme precision in order to be able to generalize it's predictions, which can be a difficult task in cloudy sky conditions. By comparison, the $k_t$ value does not show this limitation (\red{lower} panel of \autoref{im: model_target_example}), since the seasonal variance of the GHI values is already taken into account by the clear sky model. Another value that is not affected by this seasonal variance is the clearness index $K_t$, which just normalizes the GHI values by $I^{\text{extr}}$ instead of $I^{\text{clr}}$. 


While the use of $k_t$ or $K_t$ as target solves the seasonal variance problem, it fails to take into account the diurnal variance of the GHI, i.e. a sample at the end of the day (with low GHI magnitude) will be weighted the same as a sample in the middle of the day (with high GHI magnitude). \redhighlight{This could hurt the overall performance of the model, as it reduces the emphasis on middle-of-the-day samples, which are typically associated with larger prediction errors (RMSE and MAE).} To solve this, we rescale the model prediction back into the GHI estimate before computing the squared error, leading to the loss function:
\begin{equation}
\label{ktw_loss}
L_\text{MSE}^{\text{weighted}} = \frac{1}{B} \sum_{i=1}^B \left(I^\text{clr/extr}(y_i - \hat{y}_i)\right)^2 = \frac{1}{B} \sum_{i=1}^B (I_i - \hat{I}_i)^2 
\end{equation}
where
\begin{equation}
I^\text{clr/extr} = \begin{cases}
I^{\text{clr}}, &\text{if $y_i = k_{t,i}$},\\ 
I^{\text{extr}}, &\text{if $y_i = K_{t,i}$}
\end{cases}
\end{equation}
and $\hat{I}_i$ is given by \eqref{GHI_estimate}.
Note that this is equivalent to the solution presented in \cite{Paletta.etal.2024.Improving-Cross-site-Generalisability}.

The results for this experiment are summarized in \autoref{tab: model_target_results_table}, from which it is clear that modeling the raw GHI values yields the worst results by a large margin. This is especially true for $\mathcal{D}^{\text{folsom}}$, where there is almost a 50\% improvement when using other target variables. \redhighlight{However, regarding the other target variables, the performance was largely the same, even when using the weighted loss function, which suggests that weighting midday samples more heavily may not be as beneficial as was initially hypothesized. Thus, we opt to use $k_t$ as our default target variable.}



\begin{figure}
\centering
\includegraphics[width=\linewidth]{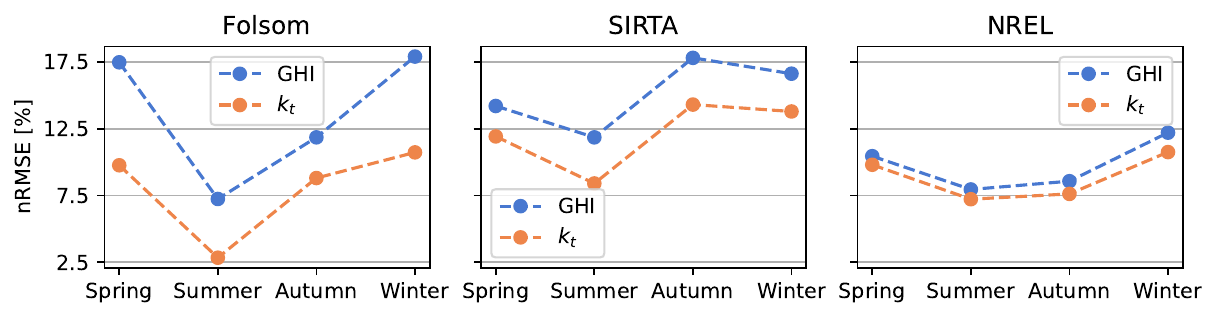}
\caption{\redhighlight{Seasonal nRMSE for the GHI and $k_t$ target variables.}} \label{im: weather_analysis}
\end{figure}

\begin{figure}
\centering
\includegraphics[width=\linewidth]{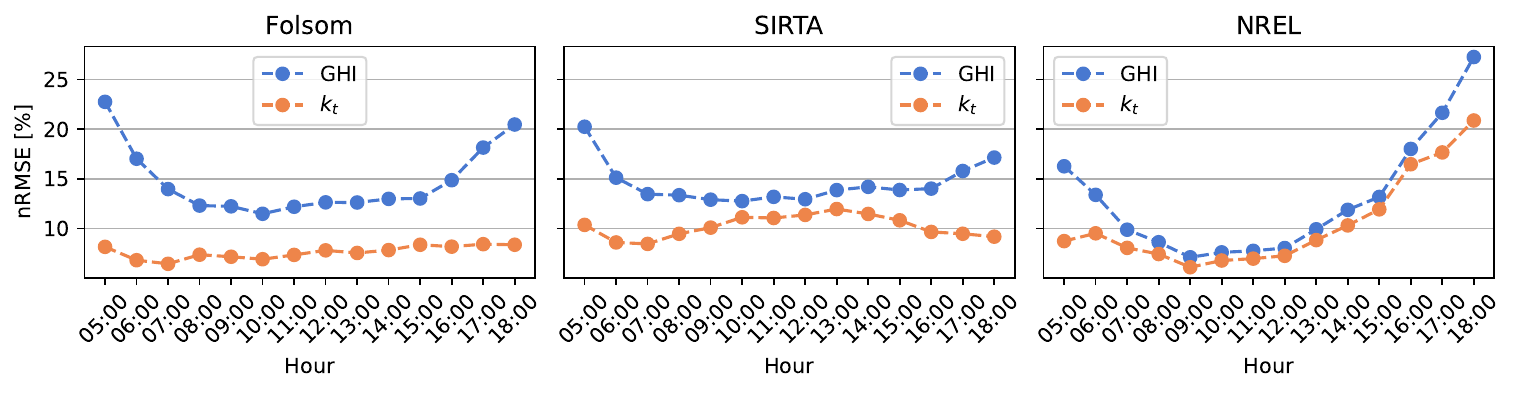}
\caption{\redhighlight{Hourly nRMSE for the GHI and $k_t$ target variables. For each hour mark, the nRMSE is computed across all samples between zero and 59 minutes.}} \label{im: hourly_analysis}
\end{figure}

\redhighlight{We also compare model performance across seasons and hours of the day, using either GHI or $k_t$ as target variable, and depict the results in} \autoref{im: weather_analysis} and \autoref{im: hourly_analysis}\redhighlight{. Since both the season and hour of day affect the magnitude of the GHI values, scale-dependent metric comparisons (such as RMSE and MAE) become more difficult. Consequently, we make use of the nRMSE metric for this analysis, due to its scale-invariant nature.}



\redhighlight{From} \autoref{im: weather_analysis}\redhighlight{ one can see that using $k_t$ over GHI as the target variable yields a consistent gain in performance across all seasons in  $\mathcal{D}^{\text{sirta}}$ and $\mathcal{D}^{\text{nrel}}$, while in $\mathcal{D}^{\text{folsom}}$ this gain was slightly larger in spring and winter. We also note that the model performance tended to be best during summer and worse in winter. Since the summer season typically contains the highest number of clear sky samples, GHI estimation becomes much easier during this season. Additionally, extended daylight hours during summer increase the number of available sky images, which may introduce a slight model bias towards this season.}

\redhighlight{Regarding the hourly results depicted in} \autoref{im: hourly_analysis}, \redhighlight{we observe that models trained with $k_t$ target consistently outperform those trained with GHI throughout the day, with particularly notable improvements around sunrise and sunset. This is significant, as these periods typically coincide with large net load ramps, where accurate irradiance predictions are critical for efficient grid management.}



\subsubsection{Camera Exposure}
\label{camera_exposure}
As shown in \autoref{im: image_data}, the images in $\mathcal{D}^{\text{sirta}}$ and $\mathcal{D}^{\text{nrel}}$ provide two different camera exposure levels, a long exposure form and a short exposure form. Although we use the long exposure form as our default value for all other experiments, the short exposure form provides more detail in the Sun disk area at the cost of providing less detail everywhere else. We hypothesize that in the \redhighlight{GHI estimation} task, predicting the irradiance value given a short exposure image is a much easier task, since the \redhighlight{main} region of interest is the area around the Sun. To test this hypothesis, we train and evaluate models using the short exposure images in $\mathcal{D}^{\text{sirta}}$ and $\mathcal{D}^{\text{nrel}}$ and compare the results to their long exposure counterparts.

\redhighlight{From the results presented in} \autoref{tab: camera_exposure_table}, \redhighlight{using the short exposure images improved the performance on both datasets, especially under cloudy sky conditions. This behavior is expected, since the long exposure images can have saturated pixels around that Sun that make it hard to discern between a Sun and a cloud pixel.} \redhighlight{Although this confirms our hypothesis that the short exposure images yield better results on the GHI estimation task, we have not tested it on the forecasting task, which is why we default to the long exposure images in our other experiments.}

\begin{table*}[t]
\centering
\caption{Test set results using the two camera exposures available in $\mathcal{D}^{\text{sirta}}$ and $\mathcal{D}^{\text{nrel}}$.}
\label{tab: camera_exposure_table}
\begin{tabular}{llllllll}
\toprule
 &  & \multicolumn{3}{c}{RMSE [W/m²]} & \multicolumn{3}{c}{MAE [W/m²]} \\
 \cmidrule(r){3-5}\cmidrule(r){6-8}
 & Exposure & All & Cloudy & Clear & All & Cloudy & Clear \\
Dataset &  &  &  &  &  &  &  \\
\midrule
SIRTA & Long & 39.68 ± 0.21 &  43.46 ± 0.22 & 10.75 ± 0.56 & 23.06 ± 0.16 & 26.30 ± 0.14 & 8.07 ± 0.42 \\
& Short & 36.46 ± 0.31 & 39.86 ± 0.34 & 11.19 ± 0.45 & 21.62 ± 0.15 & 24.58 ± 0.15 & 7.93 ± 0.31 \\
\cline{1-8}
\redhighlight{NREL} & Long & 38.75 ± 0.29 & 49.85 ± 0.37 & 10.25 ± 0.58 & 21.73 ±  0.23 & 31.56 ±  0.38 & 7.77 ±  0.41 \\
& Short & 37.34 ± 0.45 & 48.03 ± 0.61 & 9.86 ± 0.30 & 21.01 ± 0.22 & 30.52 ± 0.35 & 7.52 ± 0.32\\
\cline{1-8}
\bottomrule
\end{tabular}
\end{table*}

\begin{table*}[b]
\centering
\caption{Sun mask results for the testing set in all three datasets. Note that the approach of \cite{Papatheofanous.etal.2022.Deep-Learning-Based-Image} uses 2014 in $\mathcal{D}^{\text{folsom}}$ as the testing set, but in our approach we use 2016.}
\label{tab: sun_mask_results_table}
\begin{tabular}{llcccc}
\toprule
 &  & \multicolumn{2}{c}{RMSE [W/m²]} & \multicolumn{2}{c}{MAE [W/m²]} \\
 \cmidrule(r){3-4}\cmidrule(r){5-6}
 & Target & GHI & $k_t$ & GHI & $k_t$ \\
Dataset &  &  &  &  &  \\
\midrule
\multirow[t]{2}{*}{Folsom} & No Mask \cite{Papatheofanous.etal.2022.Deep-Learning-Based-Image} & 64.83 & - & 37.23 & - \\
& Mask \cite{Papatheofanous.etal.2022.Deep-Learning-Based-Image} & 60.31 & - & \textbf{36.02} & - \\
& No Mask [Ours] & 65.77 ± 2.22 & \textbf{37.21 ± 0.26} & 48.24 ± 2.22 & \textbf{20.58 ± 0.15} \\
& Mask [Ours] & \textbf{56.60 ± 4.95} & 39.90 ± 0.52 & 41.18 ± 5.13 & 22.79 ± 0.40 \\
\cline{1-6}
\multirow[t]{2}{*}{SIRTA} & No Mask & 50.84 ± 2.18 & \textbf{39.68 ± 0.21} & 34.69 ± 1.92 & \textbf{23.06 ± 0.16} \\
& Mask & \textbf{43.67 ± 1.51} & 40.83 ± 0.74 & \textbf{27.31 ± 1.45} & 24.04 ± 0.69 \\
\cline{1-6}
\multirow[t]{2}{*}{\redhighlight{NREL}} & No Mask & \textbf{42.43 ± 0.68} & \textbf{38.75 ± 0.29} & 25.94 ± 0.36 & \textbf{21.73 ± 0.23} \\
& Mask & 42.98 ± 0.52 & 41.62 ± 0.73 & \textbf{25.48 ± 0.40} & 22.83 ± 0.34 \\
\cline{1-6}
\bottomrule
\end{tabular}
\end{table*}

\subsubsection{Sample Interval}
\label{sample_interval}
In this experiment we change the size of the training set by increasing the sample interval from 1 to 10 minutes. Since in $\mathcal{D}^{\text{nrel}}$ the minimum sample interval is 10 minutes, we do not search for any sample interval values using this dataset. All other configurations are kept at their default values.

\begin{figure}
\centering
\includegraphics[width=\linewidth]{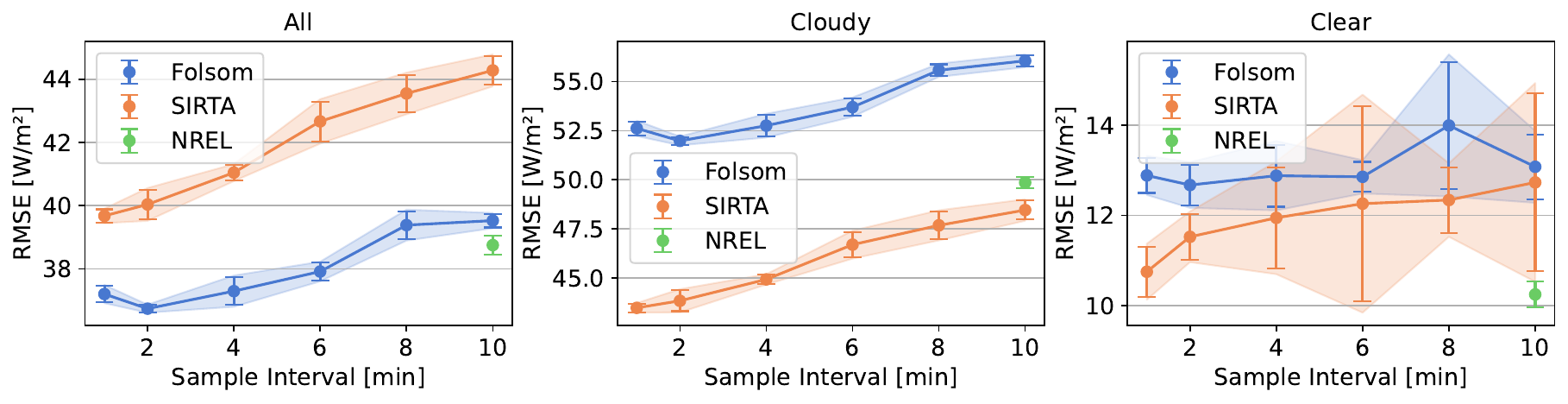}
\caption{\redhighlight{Test set RMSE for different sample intervals applied only to the training set. Only a single marker is shown for $\mathcal{D}^{\text{nrel}}$ since the minimum sample interval in this dataset is 10 minutes. The 1 minute sample interval value for $\mathcal{D}^{\text{sirta}}$ is only for the 2017 training year, since in 2018 the minimum sample value is 2 minutes.}} \label{im: sample_interval_results}
\end{figure}

\redhighlight{The results depicted in} \autoref{im: sample_interval_results} \redhighlight{show that the test error rapidly increases with larger sample intervals under cloudy sky conditions, whereas under clear sky conditions the increase is much smaller. This behavior is expected since modeling clear sky conditions is a relatively simple task, simple enough that little new information would be gained by providing the model with additional clear sky training examples. In contrast, cloudy sky conditions can exhibit high variability due to complex cloud dynamics, shapes and overirradiance effects. As a result, decreasing the number of training examples rapidly decreases the model's robustness under these conditions.}

\redhighlight{Additionally, it is also interesting to note that, when all datasets are on ``equal ground'' with a 10 minute sample interval, the behavior mentioned in} \autoref{model_architectures} \redhighlight{is even more noticeable, that is, the ``Cloudy'' test set performance in $\mathcal{D}^{\text{folsom}}$ gets significantly worse by comparison. This further validates our hypothesis that the model is mistaking the lens contamination stains for false clouds.}

\subsubsection{Sun Mask}
\label{sun_mask}
In the Sun mask experiment, we compare the model performance for 4 different combinations: 1) GHI target with no mask; 2) $k_t$ target with no mask; 3) GHI target with mask; 4) $k_t$ target with mask. In 3) and 4), we append the Sun mask described in \autoref{sun_mask_generation} as a fourth channel to the RGB input image. This implies that the first convolutional layer of the model needs to be redesigned such that it accepts a 4-channel input, which also means that the kernel weights corresponding to this additional channel need to randomly initialized, since no pre-trained weights are available. All other configurations are kept at their default values presented in \autoref{tab: configs_table}.

In \autoref{tab: sun_mask_results_table} we show the \redhighlight{GHI estimation} results for the different mask/target combinations, and compare it to \cite{Papatheofanous.etal.2022.Deep-Learning-Based-Image}. Even though they used a different testing year for $\mathcal{D}^{\text{folsom}}$, we were also able to confirm, \redhighlight{for $\mathcal{D}^{\text{folsom}}$ and $\mathcal{D}^{\text{sirta}}$, that appending a Sun mask as a fourth channel to the input improves the model performance when the GHI is its target, while doing so in $\mathcal{D}^{\text{nrel}}$ seemed to yield negligible results}. However, when the $k_t$ is the target, the Sun mask actually degrades the model performance in all datasets. We hypothesize that this is because of the fact that the Sun position in the image is less relevant when modeling the $k_t$, as discussed in \autoref{model_target}, such that appending a Sun mask would mostly add redundant information to the model, which could increase the risk of over-fitting the training set. Another possible explanation is the necessary redesign of the model to accept a 4-channel input, which forced a random weight initialization in the added weights of the first layer of the model. Since all other weights are initialized with their pre-trained values, this can disrupt the pre-learned low-level features extracted by the first layer.

\begin{figure*}
\centering
\includegraphics[width=\linewidth]{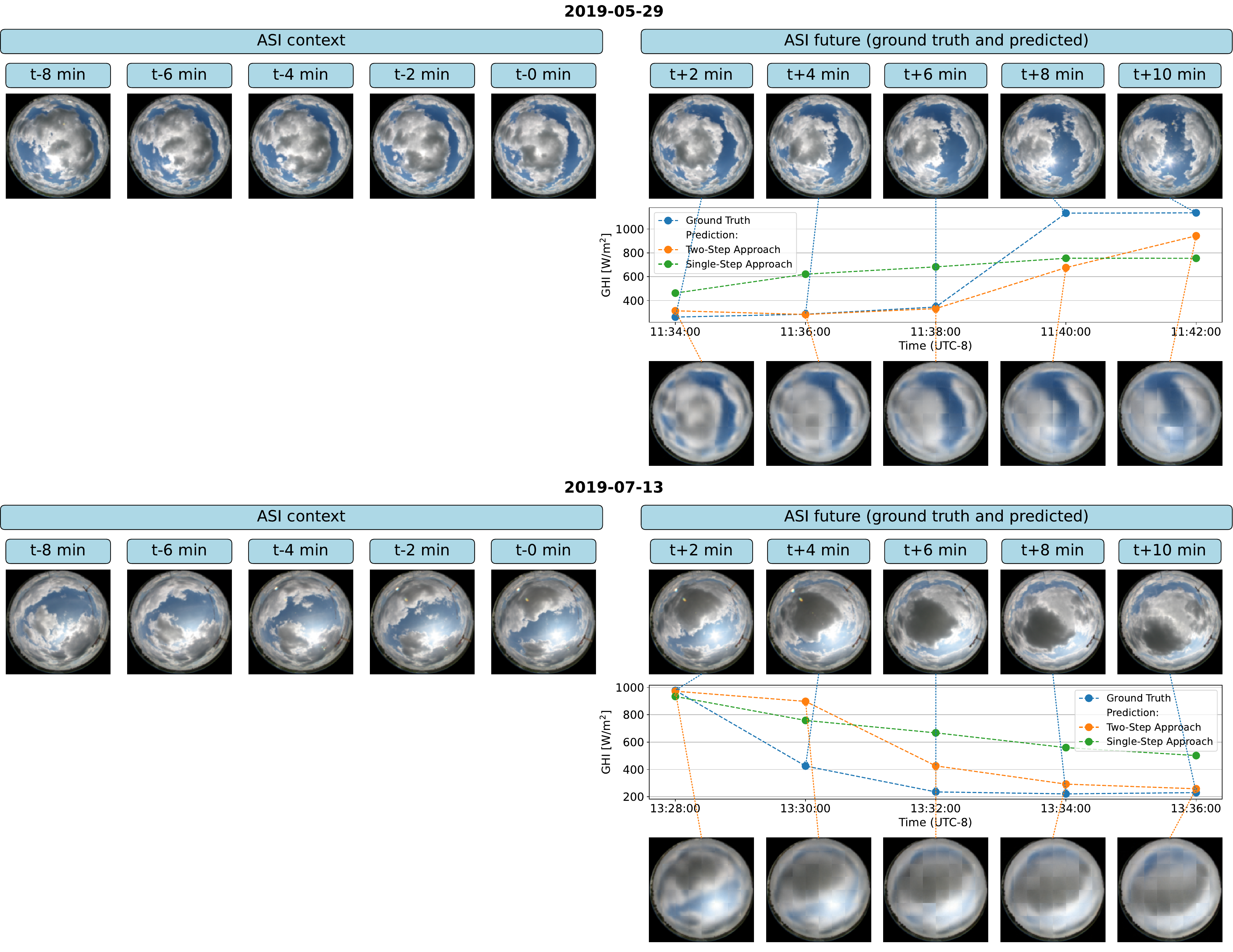}
\caption{\redhighlight{Ground truth and GHI predictions from the two-step and single-step approaches on two distinct future time windows: one where the Sun transitions from covered to visible (top panel), and the other where it transitions from visible to covered (bottom panel). In each panel, the left section displays the ASI context used as input for the forecasting model. The right section displays: ground truth future ASI (top row); ground truth and predicted future GHI values (middle row); predicted future ASI used as input for the GHI estimation model in the two-step approach (bottom row).}}\label{im: forecasting_results_img}
\end{figure*}

\begin{table*}
\centering
\caption{\redhighlight{Forecasting results for all models across all lead times. The proposed two-step approach showed significant improvements over the single-step approach, as well as the state of the art ECLIPSE} \cite{Paletta.etal.2022.ECLIPSE-Envisioning-CLoud} \redhighlight{model. A lower bound on the forecasting error of the two-step approach
is obtained by estimating the GHI using the ground truth instead of the predicted sky images.}}
\label{tab: forecasting_results_tab}
\begin{tabular}{lrrrrr}
\toprule
 & \multicolumn{5}{c}{RMSE [W/m$^2$] (Forecast Skill [\%])}  \\
 \cmidrule(r){2-6}
& 2 min & 4 min & 6 min & 8 min & 10 min \\
\midrule
Smart Persistence & 93.6 (0.0\%) & 117.4 (0.0\%) & 129.8 (0.0\%) & 137.6 (0.0\%) & 143.1 (0.0\%) \\
ECLIPSE \cite{Paletta.etal.2022.ECLIPSE-Envisioning-CLoud} & 83.8 (10.2\%) & - & 98.5 (23.6\%) & - & 109.1 (24.0\%)\\
\midrule
Single-step approach & 90.5 (3.3\%) & 102.2 (13.0\%) & 108.2 (16.7\%) & 111.1 (19.3\%) & 113.9 (20.4\%) \\
Two-step approach (Proposed) & \textbf{80.9 (13.6\%)} & \textbf{88.8 (24.3\%)} & \textbf{94.6 (27.2\%)} & \textbf{99.8 (27.5\%)} & \textbf{103.9 (27.4\%)} \\
\midrule
GHI estimation from real future ASI
& 44.1 (52.9\%) & 44.1 (62.4\%) & 44.1 (66.0\%) & 44.1 (67.9\%) & 44.1 (69.2\%) \\
\midrule
\bottomrule
\end{tabular}
\end{table*}

\subsection{Forecasting Results}
\redhighlight{We begin our forecasting analysis by comparing the predictions from the two-step and single-step approaches on two distinct time windows. The first (top panel of} \autoref{im: forecasting_results_img})\redhighlight{, where the Sun transitions from covered to visible, corresponds to the window with the highest standard deviation in \textbf{ground truth} future GHI values. The second (bottom panel of} \autoref{im: forecasting_results_img}\redhighlight{), where the Sun transitions from visible to covered, corresponds to the window with the highest standard deviation in \textbf{predicted} future GHI values. Both cases show that the single-step approach tends to avoid large errors by simply predicting a mean GHI value, which comes at the cost of failing to predict GHI ramp events. In contrast, the two-step approach is capable of identifying such events, albeit with some limitations. By decoupling future event prediction from GHI estimation, the model is able to capture the cloud dynamics more effectively, enabling it to correctly predict when the Sun will be covered and when it will be visible. Consequently, when the generated images are combined with a GHI estimation model, the two-step approach yields much more accurate and sensible predictions.}


\redhighlight{However, we note that the generated images are far from realistic, exhibiting noticeable blur and heavy artifacts along the patch borders. The blur is likely due to our choice of loss function, which was only the pixel-wise MSE, while the border artifacts likely stems from our choice of linear decoder, which independently decodes each patch embedding. Nevertheless, we believe these results provide a strong proof of concept for the two-step approach in GHI nowcasting.}

\redhighlight{In} \autoref{tab: forecasting_results_tab} \redhighlight{we compare the test set results for all forecasting models. These results further corroborate our proposed two-step approach, since it shows a significant improvement over the other models across all lead times. However, the lower bound forecasting error, which we obtain by replacing the predicted sky images with the ground truth before the GHI estimation module, indicates that there is still a large performance gap to be gained by improving the forecasting module and generating more realistic images.}

\section{Conclusion}
\label{conclusion}
\redhighlight{Due to the intermittent nature of solar PV energy, the growing integration of this energy source into the electrical grid presents several technical challenges, such as optimizing the operational cost of the grid and maintaining its frequency and voltage stability. To address this issue, recent studies have applied deep learning-based architectures and methods to forecast the short-term irradiance (i.e. nowcasting) from all-sky images. In this study, we approach the nowcasting problem using a two-step framework: first we predict the future ASI with a video prediction module; second we estimate the corresponding GHI to each predicted future image frame with an irradiance estimation module. Our primary focus is on the GHI estimation module, and we begin our analysis with a benchmark of 10 different model architectures. This benchmark revealed that the ResNet50 model was the best performer across all datasets and highlighted that model depth and residual connections can improve model performance, especially under cloudy sky conditions.}

\redhighlight{Using the ResNet50 model, we then conduct ablation experiments and uncover several key findings: 1) in the Folsom dataset, the image file names are misaligned with the irradiance timestamps, but the correct alignment can be achieved by using the image date modified metadata; 2) shifting the irradiance measurements by a small $\Delta \text{t}$ during training improves performance by mitigating the backward/forward averaging effects applied to these measurements; 3) predicting the clear-sky index $k_t$ yields substantial performance gains compared to predicting GHI directly; 4) appending a Sun mask as a fourth image channel only benefits the model performance when predicting the GHI directly; among other insights.}

\redhighlight{We conclude our study by combining our best GHI estimation model with a video prediction model, achieving state of the art performance on the SIRTA dataset and providing strong proof of concept results for this two-step approach. These results highlight the potential for significant advancements in both the video prediction and GHI estimation modules, and we hope they inspire further research using the two-step approach in the GHI nowcasting literature.}

\section*{Declaration of Competing Interest}
The authors declare that they have no competing interests.

\section*{Data Availability}
All three datasets used in this study are publicly available online. The Folsom dataset \cite{Pedro.etal.2019.Comprehensive-Dataset-Accelerated} is available at \url{https://doi.org/10.5281/zenodo.2826939}. The NREL dataset \cite{Stoffel.Andreas.1981.NREL-Solar-Radiation} is available at \url{https://doi.org/10.7799/1052221}. The SIRTA dataset \cite{Haeffelin.etal.2005.SIRTA-Ground-based-Atmospheric} can be requested by filling out a data request form, which is to be found at the SIRTA observatory website.

\section*{Acknowledgements}
The authors would like to acknowledge the SIRTA observatory for kindly providing the sky image and irradiance data used in this study. The authors would also like to acknowledge the authors of the Folsom and NREL datasets, for open-sourcing their respective datasets.

The work of D. Silva was supported in part by the Brazilian National Council for Scientific and Technological Development (CNPq-Brazil) under Grant 304619/2022-1.

The work of L. Varaschin was supported in part by Fundacao de Amparo à Pesquisa e Inovacao do Estado de Santa Catarina (FAPESC), Edital 61/2024.

This work was supported in part by computational resources funded by FAPESC under Grant 2024TR000090.

\bibliographystyle{elsarticle-num} 
\biboptions{sort&compress}
\bibliography{references}

\begin{thebibliography}{10}
\expandafter\ifx\csname url\endcsname\relax
  \def\url#1{\texttt{#1}}\fi
\expandafter\ifx\csname urlprefix\endcsname\relax\def\urlprefix{URL }\fi
\expandafter\ifx\csname href\endcsname\relax
  \def\href#1#2{#2} \def\path#1{#1}\fi

\bibitem{Kabir.etal.2018.Solar-Energy-Potential}
E.~Kabir, P.~Kumar, S.~Kumar, A.~A. Adelodun, K.-H. Kim, Solar energy:
  {{Potential}} and future prospects, Renewable and Sustainable Energy Reviews
  82 (2018) 894--900.
\newblock \href {https://doi.org/10.1016/j.rser.2017.09.094}
  {\path{doi:10.1016/j.rser.2017.09.094}}.

\bibitem{Liu.Bebic.2008.Distribution-System-Voltage}
E.~Liu, J.~Bebic, Distribution {{System Voltage Performance Analysis}} for
  {{High-Penetration Photovoltaics}}, Tech. Rep. NREL/SR-581-42298, 924648
  (Feb. 2008).
\newblock \href {https://doi.org/10.2172/924648} {\path{doi:10.2172/924648}}.

\bibitem{Kakimoto.etal.2011.Voltage-Control-Photovoltaic}
N.~Kakimoto, Q.-Y. Piao, H.~Ito, Voltage {{Control}} of {{Photovoltaic
  Generator}} in {{Combination With Series Reactor}}, IEEE Transactions on
  Sustainable Energy 2~(4) (2011) 374--382.
\newblock \href {https://doi.org/10.1109/TSTE.2011.2148181}
  {\path{doi:10.1109/TSTE.2011.2148181}}.

\bibitem{Eftekharnejad.etal.2015.Optimal-Generation-Dispatch}
S.~Eftekharnejad, G.~T. Heydt, V.~Vittal, Optimal {{Generation Dispatch With
  High Penetration}} of {{Photovoltaic Generation}}, IEEE Transactions on
  Sustainable Energy 6~(3) (2015) 1013--1020.
\newblock \href {https://doi.org/10.1109/TSTE.2014.2327122}
  {\path{doi:10.1109/TSTE.2014.2327122}}.

\bibitem{Fan.etal.2013.Probabilistic-Power-Flow}
M.~Fan, V.~Vittal, G.~T. Heydt, R.~Ayyanar, Probabilistic {{Power Flow Analysis
  With Generation Dispatch Including Photovoltaic Resources}}, IEEE
  Transactions on Power Systems 28~(2) (2013) 1797--1805.
\newblock \href {https://doi.org/10.1109/TPWRS.2012.2219886}
  {\path{doi:10.1109/TPWRS.2012.2219886}}.

\bibitem{Bird.etal.2013.Integrating-Variable-Renewable}
L.~Bird, M.~Milligan, D.~Lew, Integrating {{Variable Renewable Energy}}:
  {{Challenges}} and {{Solutions}}, Tech. Rep. NREL/TP-6A20-60451, 1097911
  (Sep. 2013).
\newblock \href {https://doi.org/10.2172/1097911} {\path{doi:10.2172/1097911}}.

\bibitem{Shah.etal.2015.Review-Key-Power}
R.~Shah, N.~Mithulananthan, R.~Bansal, V.~Ramachandaramurthy, A review of key
  power system stability challenges for large-scale {{PV}} integration,
  Renewable and Sustainable Energy Reviews 41 (2015) 1423--1436.
\newblock \href {https://doi.org/10.1016/j.rser.2014.09.027}
  {\path{doi:10.1016/j.rser.2014.09.027}}.

\bibitem{Rajagukguk.etal.2020.Review-Deep-Learning}
R.~A. Rajagukguk, R.~A.~A. Ramadhan, H.-J. Lee, A {{Review}} on {{Deep Learning
  Models}} for {{Forecasting Time Series Data}} of {{Solar Irradiance}} and
  {{Photovoltaic Power}}, Energies 13~(24) (2020) 6623.
\newblock \href {https://doi.org/10.3390/en13246623}
  {\path{doi:10.3390/en13246623}}.

\bibitem{Wen.etal.2021.Deep-Learning-Based}
H.~Wen, Y.~Du, X.~Chen, E.~Lim, H.~Wen, L.~Jiang, W.~Xiang, Deep {{Learning
  Based Multistep Solar Forecasting}} for {{PV Ramp-Rate Control Using Sky
  Images}}, IEEE Transactions on Industrial Informatics 17~(2) (2021)
  1397--1406.
\newblock \href {https://doi.org/10.1109/TII.2020.2987916}
  {\path{doi:10.1109/TII.2020.2987916}}.

\bibitem{He.etal.2015.Deep-Residual-Learning}
K.~He, X.~Zhang, S.~Ren, J.~Sun, Deep {{Residual Learning}} for {{Image
  Recognition}} (Dec. 2015).
\newblock \href {http://arxiv.org/abs/1512.03385} {\path{arXiv:1512.03385}}.

\bibitem{Yang.etal.2021.3D-CNN-Based-Sky-Image}
H.~Yang, L.~Wang, C.~Huang, X.~Luo, {{3D-CNN-Based Sky Image Feature
  Extraction}} for {{Short-Term Global Horizontal Irradiance Forecasting}},
  Water 13~(13) (2021) 1773.
\newblock \href {https://doi.org/10.3390/w13131773}
  {\path{doi:10.3390/w13131773}}.

\bibitem{Tran.etal.2015.Learning-Spatiotemporal-Features}
D.~Tran, L.~Bourdev, R.~Fergus, L.~Torresani, M.~Paluri, Learning
  {{Spatiotemporal Features}} with {{3D Convolutional Networks}} (Oct. 2015).
\newblock \href {http://arxiv.org/abs/1412.0767} {\path{arXiv:1412.0767}}.

\bibitem{Paletta.etal.2021.Benchmarking-Deep-Learning}
Q.~Paletta, G.~Arbod, J.~Lasenby, Benchmarking of deep learning irradiance
  forecasting models from sky images -- {{An}} in-depth analysis, Solar Energy
  224 (2021) 855--867.
\newblock \href {https://doi.org/10.1016/j.solener.2021.05.056}
  {\path{doi:10.1016/j.solener.2021.05.056}}.

\bibitem{Liu.etal.2023.Transformer-based-Multimodal-learning-Framework}
J.~Liu, H.~Zang, L.~Cheng, T.~Ding, Z.~Wei, G.~Sun, A {{Transformer-based}}
  multimodal-learning framework using sky images for ultra-short-term solar
  irradiance forecasting, Applied Energy 342 (2023) 121160.
\newblock \href {https://doi.org/10.1016/j.apenergy.2023.121160}
  {\path{doi:10.1016/j.apenergy.2023.121160}}.

\bibitem{Vaswani.etal.2023.Attention-All-You}
A.~Vaswani, N.~Shazeer, N.~Parmar, J.~Uszkoreit, L.~Jones, A.~N. Gomez,
  L.~Kaiser, I.~Polosukhin, Attention {{Is All You Need}} (Aug. 2023).
\newblock \href {http://arxiv.org/abs/1706.03762} {\path{arXiv:1706.03762}}.

\bibitem{Dosovitskiy.etal.2021.Image-Worth-16x16}
A.~Dosovitskiy, L.~Beyer, A.~Kolesnikov, D.~Weissenborn, X.~Zhai,
  T.~Unterthiner, M.~Dehghani, M.~Minderer, G.~Heigold, S.~Gelly, J.~Uszkoreit,
  N.~Houlsby, An {{Image}} is {{Worth}} 16x16 {{Words}}: {{Transformers}} for
  {{Image Recognition}} at {{Scale}} (Jun. 2021).
\newblock \href {http://arxiv.org/abs/2010.11929} {\path{arXiv:2010.11929}}.

\bibitem{Feng.Zhang.2020.SolarNet-Sky-Image-based}
C.~Feng, J.~Zhang, {{SolarNet}}: {{A}} sky image-based deep convolutional
  neural network for intra-hour solar forecasting, Solar Energy 204 (2020)
  71--78.
\newblock \href {https://doi.org/10.1016/j.solener.2020.03.083}
  {\path{doi:10.1016/j.solener.2020.03.083}}.

\bibitem{Simonyan.Zisserman.2015.Very-Deep-Convolutional}
K.~Simonyan, A.~Zisserman, Very {{Deep Convolutional Networks}} for
  {{Large-Scale Image Recognition}} (Apr. 2015).
\newblock \href {http://arxiv.org/abs/1409.1556} {\path{arXiv:1409.1556}}.

\bibitem{Sun.etal.2019.Short-term-Solar-Power-1}
Y.~Sun, V.~Venugopal, A.~R. Brandt, Short-term solar power forecast with deep
  learning: {{Exploring}} optimal input and output configuration, Solar Energy
  188 (2019) 730--741.
\newblock \href {https://doi.org/10.1016/j.solener.2019.06.041}
  {\path{doi:10.1016/j.solener.2019.06.041}}.

\bibitem{Nie.etal.2024.SkyGPT-Probabilistic-Ultra-short-term}
Y.~Nie, E.~Zelikman, A.~Scott, Q.~Paletta, A.~Brandt, {{SkyGPT}}:
  {{Probabilistic}} ultra-short-term solar forecasting using synthetic sky
  images from physics-constrained {{VideoGPT}}, Advances in Applied Energy 14
  (2024) 100172.
\newblock \href {https://doi.org/10.1016/j.adapen.2024.100172}
  {\path{doi:10.1016/j.adapen.2024.100172}}.

\bibitem{Ronneberger.etal.2015.U-Net-Convolutional-Networks}
O.~Ronneberger, P.~Fischer, T.~Brox, U-{{Net}}: {{Convolutional Networks}} for
  {{Biomedical Image Segmentation}} (May 2015).
\newblock \href {http://arxiv.org/abs/1505.04597} {\path{arXiv:1505.04597}}.

\bibitem{Papatheofanous.etal.2022.Deep-Learning-Based-Image}
E.~A. Papatheofanous, V.~Kalekis, G.~Venitourakis, F.~Tziolos, D.~Reisis, Deep
  {{Learning-Based Image Regression}} for {{Short-Term Solar Irradiance
  Forecasting}} on the {{Edge}}, Electronics 11~(22) (2022) 3794.
\newblock \href {https://doi.org/10.3390/electronics11223794}
  {\path{doi:10.3390/electronics11223794}}.

\bibitem{Nie.etal.2023.SKIPPD-SKy-Images}
Y.~Nie, X.~Li, A.~Scott, Y.~Sun, V.~Venugopal, A.~Brandt, {{SKIPP}}'{{D}}: {{A
  SKy Images}} and {{Photovoltaic Power Generation Dataset}} for short-term
  solar forecasting, Solar Energy 255 (2023) 171--179.
\newblock \href {https://doi.org/10.1016/j.solener.2023.03.043}
  {\path{doi:10.1016/j.solener.2023.03.043}}.

\bibitem{..Glossary-Solar-Radiation}
Glossary~of~solar~radiation~resource~terms:~{{National~Renewable~Energy~Laboratory}}.

\bibitem{Antonanzas-Torres.etal.2019.Clear-Sky-Solar}
F.~{Antonanzas-Torres}, R.~Urraca, J.~Polo, O.~{Perpi{\~n}{\'a}n-Lamigueiro},
  R.~Escobar, Clear sky solar irradiance models: {{A}} review of seventy
  models, Renewable and Sustainable Energy Reviews 107 (2019) 374--387.
\newblock \href {https://doi.org/10.1016/j.rser.2019.02.032}
  {\path{doi:10.1016/j.rser.2019.02.032}}.

\bibitem{Sun.etal.2019.Worldwide-Performance-Assessment}
X.~Sun, J.~M. Bright, C.~A. Gueymard, B.~Acord, P.~Wang, N.~A. Engerer,
  Worldwide performance assessment of 75 global clear-sky irradiance models
  using {{Principal Component Analysis}}, Renewable and Sustainable Energy
  Reviews 111 (2019) 550--570.
\newblock \href {https://doi.org/10.1016/j.rser.2019.04.006}
  {\path{doi:10.1016/j.rser.2019.04.006}}.

\bibitem{Ineichen.2008.Broadband-Simplified-Version}
P.~Ineichen, A broadband simplified version of the {{Solis}} clear sky model,
  Solar Energy 82~(8) (2008) 758--762.
\newblock \href {https://doi.org/10.1016/j.solener.2008.02.009}
  {\path{doi:10.1016/j.solener.2008.02.009}}.

\bibitem{DoNascimento.etal.2019.Extreme-Solar-Overirradiance}
L.~R. Do~Nascimento, T.~De~Souza~Viana, R.~A. Campos, R.~R{\"u}ther, Extreme
  solar overirradiance events: {{Occurrence}} and impacts on utility-scale
  photovoltaic power plants in {{Brazil}}, Solar Energy 186 (2019) 370--381.
\newblock \href {https://doi.org/10.1016/j.solener.2019.05.008}
  {\path{doi:10.1016/j.solener.2019.05.008}}.

\bibitem{Inman.etal.2013.Solar-Forecasting-Methods}
R.~H. Inman, H.~T. Pedro, C.~F. Coimbra, Solar forecasting methods for
  renewable energy integration, Progress in Energy and Combustion Science
  39~(6) (2013) 535--576.
\newblock \href {https://doi.org/10.1016/j.pecs.2013.06.002}
  {\path{doi:10.1016/j.pecs.2013.06.002}}.

\bibitem{Gueymard.etal.2019.Posteriori-Clear-sky-Identification}
C.~A. Gueymard, J.~M. Bright, D.~Lingfors, A.~Habte, M.~Sengupta, A posteriori
  clear-sky identification methods in solar irradiance time series: {{Review}}
  and preliminary validation using sky imagers, Renewable and Sustainable
  Energy Reviews 109 (2019) 412--427.
\newblock \href {https://doi.org/10.1016/j.rser.2019.04.027}
  {\path{doi:10.1016/j.rser.2019.04.027}}.

\bibitem{Reno.Hansen.2016.Identification-Periods-Clear}
M.~J. Reno, C.~W. Hansen, Identification of periods of clear sky irradiance in
  time series of {{GHI}} measurements, Renewable Energy 90 (2016) 520--531.
\newblock \href {https://doi.org/10.1016/j.renene.2015.12.031}
  {\path{doi:10.1016/j.renene.2015.12.031}}.

\bibitem{Pedro.etal.2019.Comprehensive-Dataset-Accelerated}
H.~T.~C. Pedro, D.~P. Larson, C.~F.~M. Coimbra, A comprehensive dataset for the
  accelerated development and benchmarking of solar forecasting methods,
  Journal of Renewable and Sustainable Energy 11~(3) (2019) 036102.
\newblock \href {https://doi.org/10.1063/1.5094494}
  {\path{doi:10.1063/1.5094494}}.

\bibitem{Yang.etal.2020.Probabilistic-Solar-Forecasting}
D.~Yang, D.~Van Der~Meer, J.~Munkhammar, Probabilistic solar forecasting
  benchmarks on a standardized dataset at {{Folsom}}, {{California}}, Solar
  Energy 206 (2020) 628--639.
\newblock \href {https://doi.org/10.1016/j.solener.2020.05.020}
  {\path{doi:10.1016/j.solener.2020.05.020}}.

\bibitem{Song.etal.2024.Intra-hour-Solar-Irradiance}
K.~Song, K.~Wang, S.~Wang, N.~Wang, J.~Zhang, K.~Zhang, H.~Wei, Intra-hour
  solar irradiance forecasting: {{An}} end-to-end {{Transformer-based}}
  network, in: 2024 39th {{Youth Academic Annual Conference}} of {{Chinese
  Association}} of {{Automation}} ({{YAC}}), IEEE, Dalian, China, 2024, pp.
  544--549.
\newblock \href {https://doi.org/10.1109/YAC63405.2024.10598711}
  {\path{doi:10.1109/YAC63405.2024.10598711}}.

\bibitem{Wang.etal.2023.Hybrid-Ensemble-Learning}
Z.~Wang, L.~Wang, C.~Huang, X.~Luo, A {{Hybrid Ensemble Learning Model}} for
  {{Short-Term Solar Irradiance Forecasting Using Historical Observations}} and
  {{Sky Images}}, IEEE Transactions on Industry Applications 59~(2) (2023)
  2041--2049.
\newblock \href {https://doi.org/10.1109/TIA.2022.3231842}
  {\path{doi:10.1109/TIA.2022.3231842}}.

\bibitem{Ruan.etal.2024.Use-Sky-Images}
G.~Ruan, X.~Chen, E.~G. Lim, L.~Fang, Q.~Su, L.~Jiang, Y.~Du, On the use of sky
  images for intra-hour solar forecasting benchmarking: {{Comparison}} of
  indirect and direct approaches, Solar Energy 276 (2024) 112649.
\newblock \href {https://doi.org/10.1016/j.solener.2024.112649}
  {\path{doi:10.1016/j.solener.2024.112649}}.

\bibitem{Zhang.etal.2023.Advanced-Multimodal-Fusion}
L.~Zhang, R.~Wilson, M.~Sumner, Y.~Wu, Advanced multimodal fusion method for
  very short-term solar irradiance forecasting using sky images and
  meteorological data: {{A}} gate and transformer mechanism approach, Renewable
  Energy 216 (2023) 118952.
\newblock \href {https://doi.org/10.1016/j.renene.2023.118952}
  {\path{doi:10.1016/j.renene.2023.118952}}.

\bibitem{NunesMaciel.etal.2024.Hybrid-Prediction-Method}
J.~Nunes~Maciel, J.~Javier Gimenez~Ledesma, O.~Hideo Ando~Junior, Hybrid
  prediction method of solar irradiance applied to short-term photovoltaic
  energy generation, Renewable and Sustainable Energy Reviews 192 (2024)
  114185.
\newblock \href {https://doi.org/10.1016/j.rser.2023.114185}
  {\path{doi:10.1016/j.rser.2023.114185}}.

\bibitem{Haeffelin.etal.2005.SIRTA-Ground-based-Atmospheric}
M.~Haeffelin, L.~Barth{\`e}s, O.~Bock, C.~Boitel, S.~Bony, D.~Bouniol,
  H.~Chepfer, M.~Chiriaco, J.~Cuesta, J.~Delano{\"e}, P.~Drobinski, J.-L.
  Dufresne, C.~Flamant, M.~Grall, A.~Hodzic, F.~Hourdin, F.~Lapouge,
  Y.~Lema{\^i}tre, A.~Mathieu, Y.~Morille, C.~Naud, V.~No{\"e}l, W.~O'Hirok,
  J.~Pelon, C.~Pietras, A.~Protat, B.~Romand, G.~Scialom, R.~Vautard,
  {{SIRTA}}, a ground-based atmospheric observatory for cloud and aerosol
  research, Annales Geophysicae 23~(2) (2005) 253--275.
\newblock \href {https://doi.org/10.5194/angeo-23-253-2005}
  {\path{doi:10.5194/angeo-23-253-2005}}.

\bibitem{Paletta.etal.2022.ECLIPSE-Envisioning-CLoud}
Q.~Paletta, A.~Hu, G.~Arbod, J.~Lasenby, {{ECLIPSE}}: {{Envisioning CLoud
  Induced Perturbations}} in {{Solar Energy}}, Applied Energy 326 (2022)
  119924.
\newblock \href {https://doi.org/10.1016/j.apenergy.2022.119924}
  {\path{doi:10.1016/j.apenergy.2022.119924}}.

\bibitem{Jain.etal.2024.Holistic-Lightweight-Approach}
M.~Jain, P.~Yadav, S.~Dev, Holistic and {{Lightweight Approach}} for {{Solar
  Irradiance Forecasting}}, IEEE Transactions on Geoscience and Remote Sensing
  (2024) 1--1\href {https://doi.org/10.1109/tgrs.2024.3415413}
  {\path{doi:10.1109/tgrs.2024.3415413}}.

\bibitem{Paletta.etal.2024.Improving-Cross-site-Generalisability}
Q.~Paletta, Y.~Nie, Y.-M. {Saint-Drenan}, B.~Le~Saux, Improving cross-site
  generalisability of vision-based solar forecasting models with
  physics-informed transfer learning, Energy Conversion and Management 309
  (2024) 118398.
\newblock \href {https://doi.org/10.1016/j.enconman.2024.118398}
  {\path{doi:10.1016/j.enconman.2024.118398}}.

\bibitem{Nie.etal.2024.Sky-Image-based-Solar}
Y.~Nie, Q.~Paletta, A.~Scott, L.~M. Pomares, G.~Arbod, S.~Sgouridis,
  J.~Lasenby, A.~Brandt, Sky image-based solar forecasting using deep learning
  with heterogeneous multi-location data: {{Dataset}} fusion versus transfer
  learning, Applied Energy 369 (2024) 123467.
\newblock \href {https://doi.org/10.1016/j.apenergy.2024.123467}
  {\path{doi:10.1016/j.apenergy.2024.123467}}.

\bibitem{Insaf.etal.2021.Global-Horizontal-Irradiance}
I.~M. Insaf, H.~M. K.~D. Wickramathilaka, M.~A.~N. Upendra, G.~M. R.~I.
  Godaliyadda, M.~P.~B. Ekanayake, H.~M. V.~R. Herath, D.~M. L.~H. Dissawa,
  J.~B. Ekanayake, Global {{Horizontal Irradiance Modeling}} from {{Sky Images
  Using ResNet Architectures}}, in: 2021 {{IEEE}} 16th {{International
  Conference}} on {{Industrial}} and {{Information Systems}} ({{ICIIS}}), IEEE,
  Kandy, Sri Lanka, 2021, pp. 239--244.
\newblock \href {https://doi.org/10.1109/ICIIS53135.2021.9660664}
  {\path{doi:10.1109/ICIIS53135.2021.9660664}}.

\bibitem{AlAsmar.etal.2021.Improvement-Solar-Irradiance}
L.~Al~Asmar, L.~{Musson-Genon}, E.~Dupont, J.-C. Dupont, K.~Sartelet,
  Improvement of solar irradiance modelling during cloudy-sky days using
  measurements, Solar Energy 230 (2021) 1175--1188.
\newblock \href {https://doi.org/10.1016/j.solener.2021.10.084}
  {\path{doi:10.1016/j.solener.2021.10.084}}.

\bibitem{Stoffel.Andreas.1981.NREL-Solar-Radiation}
T.~Stoffel, A.~Andreas, {{NREL Solar Radiation Research Laboratory}}
  ({{SRRL}}): {{Baseline Measurement System}} ({{BMS}}); {{Golden}},
  {{Colorado}} ({{Data}}) (1981).
\newblock \href {https://doi.org/10.7799/1052221} {\path{doi:10.7799/1052221}}.

\bibitem{Jonathan.etal.2024.Radiant-Shift-Attention-embedded}
A.~L. Jonathan, D.~Cai, C.~C. Ukwuoma, N.~J.~J. Nkou, Q.~Huang, O.~Bamisile, A
  radiant shift: {{Attention-embedded CNNs}} for accurate solar irradiance
  forecasting and prediction from sky images, Renewable Energy 234 (2024)
  121133.
\newblock \href {https://doi.org/10.1016/j.renene.2024.121133}
  {\path{doi:10.1016/j.renene.2024.121133}}.

\bibitem{Gao.Liu.2022.Short-term-Solar-Irradiance}
H.~Gao, M.~Liu, Short-term {{Solar Irradiance Prediction}} from {{Sky Images}}
  with a {{Clear Sky Model}}, in: 2022 {{IEEE}}/{{CVF Winter Conference}} on
  {{Applications}} of {{Computer Vision}} ({{WACV}}), IEEE, Waikoloa, HI, USA,
  2022, pp. 3074--3082.
\newblock \href {https://doi.org/10.1109/WACV51458.2022.00313}
  {\path{doi:10.1109/WACV51458.2022.00313}}.

\bibitem{Zuo.etal.2022.Ten-minute-Prediction-Solar}
H.-M. Zuo, J.~Qiu, Y.-H. Jia, Q.~Wang, F.-F. Li, Ten-minute prediction of solar
  irradiance based on cloud detection and a long short-term memory ({{LSTM}})
  model, Energy Reports 8 (2022) 5146--5157.
\newblock \href {https://doi.org/10.1016/j.egyr.2022.03.182}
  {\path{doi:10.1016/j.egyr.2022.03.182}}.

\bibitem{Feng.etal.2022.Convolutional-Neural-Networks}
C.~Feng, J.~Zhang, W.~Zhang, B.-M. Hodge, Convolutional neural networks for
  intra-hour solar forecasting based on sky image sequences, Applied Energy 310
  (2022) 118438.
\newblock \href {https://doi.org/10.1016/j.apenergy.2021.118438}
  {\path{doi:10.1016/j.apenergy.2021.118438}}.

\bibitem{Parmar.etal.2022.Aliased-Resizing-Surprising}
G.~Parmar, R.~Zhang, J.-Y. Zhu, On {{Aliased Resizing}} and {{Surprising
  Subtleties}} in {{GAN Evaluation}} (Jan. 2022).
\newblock \href {http://arxiv.org/abs/2104.11222} {\path{arXiv:2104.11222}}.

\bibitem{fisheye_lens}
F.~Bettonvil, Fisheye lenses, WGN, Journal of the International Meteor
  Organization 33 (2005) 9--14.

\bibitem{Izmailov.etal.2019.Averaging-Weights-Leads}
P.~Izmailov, D.~Podoprikhin, T.~Garipov, D.~Vetrov, A.~G. Wilson, Averaging
  {{Weights Leads}} to {{Wider Optima}} and {{Better Generalization}} (Feb.
  2019).
\newblock \href {http://arxiv.org/abs/1803.05407} {\path{arXiv:1803.05407}}.

\bibitem{Tan.Le.2021.EfficientNetV2-Smaller-Models}
M.~Tan, Q.~V. Le, {{EfficientNetV2}}: {{Smaller Models}} and {{Faster
  Training}} (Jun. 2021).
\newblock \href {http://arxiv.org/abs/2104.00298} {\path{arXiv:2104.00298}}.

\bibitem{Radosavovic.etal.2020.Designing-Network-Design}
I.~Radosavovic, R.~P. Kosaraju, R.~Girshick, K.~He, P.~Doll{\'a}r, Designing
  {{Network Design Spaces}} (Mar. 2020).
\newblock \href {http://arxiv.org/abs/2003.13678} {\path{arXiv:2003.13678}}.

\bibitem{Sandler.etal.2019.MobileNetV2-Inverted-Residuals}
M.~Sandler, A.~Howard, M.~Zhu, A.~Zhmoginov, L.-C. Chen, {{MobileNetV2}}:
  {{Inverted Residuals}} and {{Linear Bottlenecks}} (Mar. 2019).
\newblock \href {http://arxiv.org/abs/1801.04381} {\path{arXiv:1801.04381}}.

\bibitem{Frazier.2018.Tutorial-Bayesian-Optimization}
P.~I. Frazier, A {{Tutorial}} on {{Bayesian Optimization}} (Jul. 2018).
\newblock \href {http://arxiv.org/abs/1807.02811} {\path{arXiv:1807.02811}}.

\bibitem{Kingma.Ba.2017.Adam-Method-Stochastic}
D.~P. Kingma, J.~Ba, Adam: {{A Method}} for {{Stochastic Optimization}} (Jan.
  2017).
\newblock \href {http://arxiv.org/abs/1412.6980} {\path{arXiv:1412.6980}}.

\bibitem{Loshchilov.Hutter.2019.Decoupled-Weight-Decay}
I.~Loshchilov, F.~Hutter, Decoupled {{Weight Decay Regularization}} (Jan.
  2019).
\newblock \href {http://arxiv.org/abs/1711.05101} {\path{arXiv:1711.05101}}.

\bibitem{Zhang.etal.2015.Suite-Metrics-Assessing}
J.~Zhang, A.~Florita, B.-M. Hodge, S.~Lu, H.~F. Hamann, V.~Banunarayanan, A.~M.
  Brockway, A suite of metrics for assessing the performance of solar power
  forecasting, Solar Energy 111 (2015) 157--175.
\newblock \href {https://doi.org/10.1016/j.solener.2014.10.016}
  {\path{doi:10.1016/j.solener.2014.10.016}}.

\bibitem{Frias-Paredes.etal.2016.Introducing-Temporal-Distortion}
L.~{Fr{\'i}as-Paredes}, F.~Mallor, T.~Le{\'o}n, M.~{Gast{\'o}n-Romeo},
  Introducing the {{Temporal Distortion Index}} to perform a bidimensional
  analysis of renewable energy forecast, Energy 94 (2016) 180--194.
\newblock \href {https://doi.org/10.1016/j.energy.2015.10.093}
  {\path{doi:10.1016/j.energy.2015.10.093}}.

\bibitem{Vallance.etal.2017.Standardized-Procedure-Assess}
L.~Vallance, B.~Charbonnier, N.~Paul, S.~Dubost, P.~Blanc, Towards a
  standardized procedure to assess solar forecast accuracy: {{A}} new ramp and
  time alignment metric, Solar Energy 150 (2017) 408--422.
\newblock \href {https://doi.org/10.1016/j.solener.2017.04.064}
  {\path{doi:10.1016/j.solener.2017.04.064}}.

\bibitem{tang2025predformer}
Y.~Tang, L.~Qi, F.~Xie, X.~Li, C.~Ma, M.-H. Yang,
  \href{https://openreview.net/forum?id=avNVrQ8D2v}{Predformer: Transformers
  are effective spatial-temporal predictive learners} (2025).
\newline\urlprefix\url{https://openreview.net/forum?id=avNVrQ8D2v}

\bibitem{DBLP:journals/corr/abs-1810-04805}
J.~Devlin, M.~Chang, K.~Lee, K.~Toutanova,
  \href{http://arxiv.org/abs/1810.04805}{{BERT:} pre-training of deep
  bidirectional transformers for language understanding}, CoRR abs/1810.04805
  (2018).
\newblock \href {http://arxiv.org/abs/1810.04805} {\path{arXiv:1810.04805}}.
\newline\urlprefix\url{http://arxiv.org/abs/1810.04805}

\end{thebibliography}






\end{document}